\documentclass[times,sort&compress,3p]{elsarticle}
\journal{Journal of Multivariate Analysis}
\usepackage[labelfont=bf]{caption}

\usepackage{amsfonts,amssymb,amsthm,booktabs,color,epsfig,graphicx,hyperref,url}

\usepackage{psfrag,epsf}
\usepackage{enumerate}
\usepackage{natbib}
\usepackage{url} % not crucial - just used below for the URL 
\usepackage{float}
\usepackage{bm}
\usepackage{makecell}
\usepackage{dsfont}
\usepackage{mathtools}
\usepackage{subcaption}

\usepackage{amsmath}

\DeclareMathOperator*{\argmin}{arg\,min}

\usepackage{bbold}
\usepackage{algorithm}
\usepackage{algorithmic}

\usepackage{multirow}
\usepackage{mathrsfs} 
\hypersetup{
     colorlinks = true,
     linkcolor=blue,
            urlcolor=blue,
            citecolor=blue
}
\usepackage{xr}

\usepackage{lipsum}

\theoremstyle{plain}% Theorem-like structures provided by amsthm.sty

\newtheorem{theorem}{Theorem}
\newtheorem{proposition}{Proposition}
\newtheorem{corollary}{Corollary}

\newtheorem{lemma}{Lemma}

\newtheorem{exmp}{Example}[section]
\AtEndDocument{\refstepcounter{theorem}\label{finalthm}}
\AtEndDocument{\refstepcounter{proposition}\label{finalprop}}
  \let\code=\texttt
\begin{document}

\begin{frontmatter}

\title{Generalized Score Matching}

  \author[1]{Jiazhen Xu\corref{mycorrespondingauthor}}
  \author[1]{Janice L. Scealy}
  \author[1]{Andrew T. A. Wood}
  \author[1]{Tao Zou}
  \address[1]{Research School of Finance, Actuarial Studies and Statistics, Australian National University, Canberra ACT 2601, Australia}

\cortext[mycorrespondingauthor]{Corresponding author. Email address: \url{jiazhen.xu@anu.edu.au}}

\begin{abstract}
Score matching is an estimation procedure that has been developed for statistical models whose probability density function is known up to proportionality but whose normalizing constant is intractable, so that maximum likelihood is difficult or impossible to implement. To date, applications of score matching have focused more on continuous IID models. Motivated by various data modelling problems, this article proposes a unified asymptotic theory of generalized score matching developed under the independence assumption, covering both continuous and discrete response data, thereby giving a sound basis for score-matching-based inference. Real data analyses and simulation studies provide convincing evidence of strong practical performance of the proposed methods.
\end{abstract}


\end{frontmatter}

\section{Introduction\label{sec:1}}
Many statistical models have a probability density function that is known up to proportionality but whose normalizing constant is intractable. For such models, maximum likelihood estimation is at best computationally challenging and, at worst, not feasible to compute. Several methods of approximating intractable normalizing constants have been studied (\cite{brooks2011handbook,huber2015approximation}). However, insufficiently accurate approximation of normalizing constants will introduce inaccuracy in estimation which persists in large samples. To tackle this issue, score matching and its extensions have been developed (\cite{hyvarinen2005estimation,hyvarinen2007some,vincent2011connection,lyu2012interpretation,song2020sliced,scealy2022score}) to avoid the explicit computation of the normalizing constant. Score matching is a powerful method for performing parameter estimation in previously intractable models.

The idea underlying score matching is to choose the unknown parameter vector to minimize the Fisher divergence (\cite{johnson2004information,dasgupta2008asymptotic}) between the parametric model density and the true density. At first glance, it does not appear to be possible to minimise this Fisher divergence because it depends on the unknown true density. However, the key insight of \cite{hyvarinen2005estimation} is that minimising the Fisher divergence between the parametric density and true density is equivalent to minimising a function for which a fully explicit unbiased estimator exists, where calculation of the latter does not require any knowledge of the true probability density function. It should be emphasized that the score function as used in score matching is the gradient of the log density with respect to the data vector, as opposed to the classical score statistic, which is the derivative of the log density with respect to the parameter vector. 
%\cite{hyvarinen2005estimation} showed that, under mild regularity conditions, the score matching objective function admits an unbiased estimator that does not require any knowledge of true probability density model. 

Score matching approaches have been applied in many areas including: graphical models (\cite{yu2016statistical,yu2019generalized,yu2020simultaneous}); directional data modelling (\cite{mardia2016score}) and compositional data modelling (\cite{yu2021interaction,scealy2022score}), where the latter paper includes an application to microbiome data; generative modelling, including applications to image analysis (\cite{song2020score,song2020improved,vahdat2021score}); and generative adversarial networks (\cite{pang2020efficient}). A robust version of score matching, using Windham's method of robustification (\cite{windham1995robustifying}), has been applied in \cite{scealy2024robust} to compositional data.

%\subsection{Main Contributions of the Article}

In this article we adopt a broader definition of an ordinal response variable than is customary, allowing it to include not only ordered categorical variables but also discrete numerical variables that inherit their ordering from $\mathbb{R}$, such as count data. Also, a discretely-distributed random vector is said to be ordinal if each of its components is ordinal.

To date, applications of score matching have focused more on continuous IID (independent and identically distributed) models. Motivated by various data modelling problems for which the continuity and/or the IID assumption are not appropriate, this article first proposes (i) a novel extension of score matching to univariate and multivariate ordinal data. We then develop (ii) a unified asymptotic framework for generalized score matching estimators and related hypothesis testing under the independence assumption. In addition, we use (iii) score-matching-based inference to test for independence in certain data dependent models, specifically auto models of exponential family type (\cite{besag1974spatial}).

We briefly explain how (i)-(iii) go beyond what is currently in the literature. 
Regarding (i), the original form of score matching in \cite{hyvarinen2005estimation} is valid only for models which possess a differentiable density function. Some variants of score matching have been proposed for multivariate categorical data, notably by \cite{hyvarinen2007some}
and \cite{lyu2012interpretation}.
\cite{hyvarinen2007some} proposed a variant of score matching, called ratio matching, to deal with multivariate binary data. Ratio matching is based on minimizing the expected squared distance of the ratios of certain probabilities given by the model and the corresponding ratios in the observations. \cite{lyu2012interpretation} developed a generalization of score matching which replaces the gradient operator used in score matching for continuous data by a general linear operator. \cite{lyu2012interpretation} then focused on what is referred to as the marginalization operator and applies it to multivariate categorical data. The approaches of both \cite{hyvarinen2007some} and \cite{lyu2012interpretation} are ultimately based on a comparison of conditional distributions derived from the multivariate structure of the observations, which are applicable for some data domains of finite cardinality. Moreover, in the case of univariate discrete observations, neither of the approaches in \cite{hyvarinen2007some} and \cite{lyu2012interpretation} resolves the problem of the intractable normalizing constant, due to there being no multivariate structure to exploit. In contrast, our approach compares conditional distributions related to the neighbourhood structure on the sample space, in turn determined by the ordering of the support of the relevant ordinal distribution; this construction works equally well for univariate and multivariate ordinal distributions without any requirements on data domain cardinality. Additionally, after this paper was made public in early 2022 on arXiv, several subsequent works have employed similar ideas on comparing conditional distributions related to the neighbourhood structure on the sample space. For instance, \cite{meng2022concrete} explored applications of essentially the same method in image denosing while \cite{matsubara2023generalized}, which referenced our 2022 arXiv version of this paper, utilized our approach in Bayesian variational inference for both univariate and multivariate count data. These studies provide compelling evidence for the effectiveness and versatility of our method.

Regarding (ii), in many real-life situations, the specification of the statistical models may further depend on other extraneous factors available to us in the form of explanatory variables or covariates. Examples of such regression modelling can be found in genetic studies (\cite{yin2011sparse,cai2013covariate,cheng2014sparse}), computer vision (\cite{gustafsson2020energy}) and network data analysis (\cite{yuan2021community,zhao2021dimension}). In the literature, score matching has so far only been developed for random design models in which the joint distribution of the response and the covariate vector is assumed to be IID. However, in many studies it is more appropriate to work with fixed design regression models, and of course the IID assumption does not hold in the traditional fixed design setting. Hence there is motivation for deriving asymptotic properties of generalized score matching estimators when observations are not required to be identically distributed. We provide the following: a unified and rigorous treatment for the continuous response and ordinal response cases under the independence assumption; a convenient form for the asymptotic distribution of the change-in-score matching statistic (cf. the log-likelihood ratio statistic) considered in Theorem \ref{thm::sm diff test} (although it is surely well known, we are not aware of a convenient reference for this result); and a numerical comparison of the performance of the score-matching-based Wald statistic and the change-in-score-matching statistic.

Regarding (iii), there are many exponential family auto models, for both continuous and discrete data, where score matching has the potential to be very useful. Auto models are developed based on the specification of conditional distributions. Previous works such as \cite{besag1974spatial,kaiser2000construction,ellis2006entropy} and \cite{tansey2015vector} focused on exponential family auto models where their conditional distributions belong to exponential family and one of the most widely used exponential family auto models is the Ising model (\cite{ising1924beitrag}). Many exponential family auto models have intractable normalizing constants and, to avoid this problem, composite likelihood has been used in the literature (see \cite{varin2011overview}). However, score matching has two advantages over composite likelihood in the auto exponential family setting. First, score matching yields closed form estimates which are easily computed, whereas composite likelihood estimators are usually not available in closed form. Second, score matching takes the dependence structure of the model fully into account, whereas composite likelihood usually fails to model the dependence structure correctly. In this paper we focus on a novel von Mises-Fisher (vMF) auto model and apply it to a geochemical dataset consisting of compositional data vectors (vectors whose components are non-negative and sum to 1), distributed at different spatial locations. Each compositional vector in this geochemical survey measures the concentration of different chemical elements based on the collection of sediment or rock samples. We make use of the square-root transformation (see \cite{stephens1982use,scealy2011regression}) to map compositional data vectors to points on a sphere. The vMF auto model provides a way to capture spatial dependence between directions (i.e. unit vectors) at different spatial locations. Although the model is quite complex and the normalizing constant is intractable, score matching is feasible to implement and details are provided. A test of spatial independence based on score matching estimators is also provided.

The rest of this article is organized as follows. Section \ref{sect::sm review and extension} reviews and extends score matching for continuous data in Euclidean space and Riemannian manifolds and discusses a model for spatially dependent directional data. Section \ref{sect::sm discrete} develops a novel generalized score matching method for ordinal data, in which we show that our proposed methodology is a tractable estimation method and inherits the benefits of the original score matching approach for continuous data. Section 
\ref{Section_4} provides a unified asymptotic theory for score matching estimators for both continuous and ordinal independent data and score-matching-based inference is also developed in this section. A doctoral publication example and Monte Carlo studies are presented in Section \ref{sect::total numerical study}; the results indicate that our estimators perform well. In Section \ref{sect::sm dependent} we propose a novel auto model to capture the spatial dependence in geochemical data and a Wald test based on score matching has been developed to test for the presence of spatial dependence. The proofs of the theorems and corollaries are given in Section \ref{sect::tech details} while additional numerical results are given in the Supplementary Material.

%Modeling results for a Spain geochemical survey is also presented in Section \ref{sect::sm dependent}.

\section{Score Matching for Continuous Data}\label{sect::sm review and extension}
In this section, we present score matching for continuous data in a general setting that permits departures from independence and/or the identically distributed assumption. \cite{hyvarinen2005estimation} presented a general framework for score matching which allows for the possibility of going beyond the IID assumption but, so far as we are aware, more general settings have not been pursued in the literature to date. We first review and then extend score matching for Euclidean space in Section \ref{sub sect::sm eucli}; then we discuss the Riemannian manifold case in Section \ref{sect::sm manifold}; and finally we consider the von Mises-Fisher auto model in Section \ref{Section_2.3}.

\subsection{Score Matching in Euclidean Space}\label{sub sect::sm eucli}

Suppose we have observations $\bm{y}_1,\ldots,\bm{y}_n\in \mathbb{R}^d$ from an unknown joint probability density function (pdf) $q(\bm{y}_1,\ldots,\bm{y}_n)$. Further assume that we have a parametrized model $p(\bm{y}_1,\ldots,\bm{y}_n|\bm{\theta})$, where $\bm{\theta}\in\mathbb{R}^p$ is a vector of unknown parameters. When this probabilistic model has an intractable normalizing constant, we can use score matching to avoid the explicit computation of the normalizing constant. This probabilistic model has the form of
\[
p(\bm{y}_1,\ldots,\bm{y}_n|\bm{\theta})=\frac{1}{Z(\bm{\theta})}\widetilde{p}(\bm{y}_1,\ldots,\bm{y}_n|\bm{\theta}),
\]
where $Z(\bm{\theta})$ is the intractable normalizing constant and $\widetilde{p}(\bm{y}_1,\ldots,\bm{y}_n|\bm{\theta})$ is the unnormalized pdf. Score matching for continuous data is based on the Fisher divergence for the unknown joint distribution and the parametrized model (\cite{johnson2004information,dasgupta2008asymptotic}). The basic score matching objective function $D_{\rm SM}(q_*,p_*)$ for $q_*=q(\bm{y}_1,\ldots,\bm{y}_n)$, the true density, and $p_*=p(\bm{y}_1,\ldots,\bm{y}_n|\bm{\theta})$, the model density, is given by
\begin{align}\label{sm population obj func}
  D_{\rm SM}(q_*,p_*)=&\frac{1}{n}{\rm E} \left( \|\nabla\log q(\bm{y}_1,\ldots,\bm{y}_n)-\nabla\log p(\bm{y}_1,\ldots,\bm{y}_n|\bm{\theta})\|^2 \right),
\end{align}
where $\nabla$ is the gradient operator with respect to $\bm{y}=(\bm{y}_1^\top, \ldots , \bm{y}_n^\top)^\top$ and the expectation is taken with respect to $q_*$. Under some mild conditions, $D_{\rm SM}(q_*,p_*)$ can be decomposed as $D_{\rm SM}(q_*,p_*)=g(q_*)+d_{\rm SM}(\bm{\theta})$, where 
\begin{align*}
  d_{\rm SM}(\bm{\theta})=\frac{1}{n}  {\rm E}  \left[2\sum_{i=1}^n \sum_{j=1}^d\frac{\partial^2}{\partial y_{ij}^2}\log p(\bm{y}_1,\ldots,\bm{y}_n|\bm{\theta})+\sum_{i=1}^n\sum_{j=1}^d\left\{\frac{\partial}{\partial y_{ij}}\log p(\bm{y}_1,\ldots,\bm{y}_n|\bm{\theta})\right\}^2\right],
\end{align*}
and $g(q_*)$ is a constant depending on $q_*$ but not on $\bm{\theta}$. An empirical estimator of the population function, $d_{\rm SM}(\bm{\theta})$, is given by
\begin{align*}
  \hat{d}_{\rm SM}(\bm{\theta})=\frac{1}{n}\left[2\sum_{i=1}^n \sum_{j=1}^d\frac{\partial^2}{\partial y_{ij}^2}\log p(\bm{y}_1,\ldots,\bm{y}_n|\bm{\theta})+\sum_{i=1}^n \sum_{j=1}^d\left\{\frac{\partial}{\partial y_{ij}}\log p(\bm{y}_1,\ldots,\bm{y}_n|\bm{\theta})\right\}^2\right].
\end{align*}
The score matching estimator for $\bm{\theta}$ is then obtained by
\begin{align}\label{sm est def}
  \hat{\bm{\theta}}=\argmin_{\bm{\theta}}\hat{d}_{\rm SM}(\bm{\theta}).
\end{align}

The score matching framework described above encompasses a general non-IID data framework. When the data are independent, the joint distribution is equal to the product of the marginal distributions, i.e., $q(\bm{y}_1,\ldots,\bm{y}_n)=\prod_{i=1}^n q_i(\bm{y}_i)$ where $q_i(\bm{y}_i)$ possibly depends on the index $i$. In this case, the score matching objective function (\ref{sm population obj func}) simplifies to
\begin{align}\label{sm population obj func indept}
  D_{\rm SM}(q_*,p_*)=\frac{1}{n}\sum_{i=1}^n{\rm E} \left( \|\nabla_i\log q_i(\bm{y}_i)-\nabla_i\log p_i(\bm{y}_i|\bm{\theta})\|^2 \right),
\end{align}
where $\nabla_i = \partial /\partial\bm{y}_i$ and $\hat{d}_{\rm SM}(\bm{\theta})$ is given by
\begin{align}\label{emprical sm cts}
  \hat{d}_{\rm SM}(\bm{\theta})=\frac{1}{n}\sum_{i=1}^n \rho_i^{\rm SM}(\bm{y}_i|\bm{\theta}),
\end{align}
where
\begin{align}\label{rho sm}
  \rho_i^{\rm SM}(\bm{y}_i|\bm{\theta})=2\sum_{j=1}^d\frac{\partial^2}{\partial y_{ij}^2}\log p(\bm{y}_i|\bm{\theta})+\sum_{j=1}^d\left\{\frac{\partial}{\partial y_{ij}}\log p(\bm{y}_i|\bm{\theta})\right\}^2.
\end{align}

\subsection{Score Matching for Riemannian Manifolds}\label{sect::sm manifold}
\cite{mardia2016score} defined consistent score matching estimators for the von Mises-Fisher, Bingham and Kent distributions under the independence assumption. The method is similar to the score matching defined by (\ref{sm population obj func indept}), but adapted to handle estimation on a Riemannian manifold. It is assumed that the Riemannian manifold under consideration (including the product manifold considered below) is embedded in an ambient Euclidean space. Additionally, geochemical data motivates us to consider joint distributions in score matching and to further extend score matching to deal with a product of Riemannian manifolds. First note that (\ref{sm population obj func}) can be represented via the inner product on gradient vectors as
\begin{align}\label{eq::sm manifold represent}
D_{\rm SM}(q_*,p_*)=\frac{1}{n}{\rm E} \left( \left\langle\log\frac{q(\bm{y}_1,\ldots,\bm{y}_n)}{  p(\bm{y}_1,\ldots,\bm{y}_n|\bm{\theta})},\log\frac{ q(\bm{y}_1,\ldots,\bm{y}_n)}{ p(\bm{y}_1,\ldots,\bm{y}_n|\bm{\theta})}\right\rangle \right),
\end{align}
where the inner product $\langle \bm{u}(\bm{z}),\bm{v}(\bm{z})\rangle$ for real-valued functions on the Cartesian product of manifolds $\mathcal{M}_1\times\ldots\times\mathcal{M}_n$ is defined by
$
\langle \bm{u}(\bm{z}),\bm{v}(\bm{z})\rangle=(\nabla\widetilde{ \bm{u}}(\bm{z}))^\top\bm{P}(\nabla\widetilde{\bm{v}}(\bm{z})),
$
with $\bm{P}$ being the orthogonal projection matrix onto the tangent hyperplane at $\bm{z}$. The functions $\bm{u}(\bm{z})$ and $\bm{v}(\bm{z})$ are extended to functions $\widetilde{ \bm{u}}(\bm{z})$ and $\widetilde{\bm{v}}(\bm{z})$, $\bm{z}\in\mathcal{N}$, where $\mathcal{N}$ is a neighbourhood of $\mathcal{M}_1\times\ldots\times\mathcal{M}_n$ within the ambient Euclidean space. The expectation in (\ref{eq::sm manifold represent}) is taken on the product of manifolds. Based on \cite{mardia2016score}, who applied Stokes' Theorem, we obtain the empirical score matching objective function given by
\[
\hat{d}_{\rm SM}(\bm{\theta})=\frac{1}{n}\left[ (\nabla\log p(\bm{y}_1,\ldots,\bm{y}_n|\bm{\theta}))^\top\bm{P}\nabla\log p(\bm{y}_1,\ldots,\bm{y}_n|\bm{\theta})+2\Delta_{M} \log p(\bm{y}_1,\ldots,\bm{y}_n|\bm{\theta})  \right],
\]
where $\Delta_M$ is the Laplace-Beltrami operator and $\Delta_M \bm{u}(\bm{z})={\rm tr}(\bm{P}\nabla^\top (\bm{P}\nabla\widetilde{ \bm{u}}(\bm{z})))$.

\subsection{The von Mises-Fisher Auto Model}\label{Section_2.3}

Inspired by the construction of the exponential family auto models, we define our novel auto model for spherical data by setting the conditional distribution to be the von Mises–Fisher distribution. For observations $\bm{y}_1,\ldots,\bm{y}_n\in\mathcal{S}^{d-1}$, our proposed vMF auto model is given by
\begin{equation}
p(\bm{y}_1,\ldots,\bm{y}_n|\bm{\beta},\xi)\propto \exp\left\{ \sum_{i=1}^n\bm{\beta}^\top \bm{y}_i+\xi\sum_{i=1}^n\sum_{k\in N(i)} \bm{y}_i^\top \bm{y}_k\right\},
\label{vMF_auto_model}
\end{equation}
where the sets $\{N(i):i=1\ldots,n\}$ specify the neighborhood structure of the observations. Note that our proposed auto model is a special case of the exponential family auto models, and thus can be represented as
\begin{align*}
  p(\bm{y}_1,\ldots,\bm{y}_n|\bm{\theta}) \propto\exp\left\{  \sum_{l=1}^{d+1}\theta_l t_l(\bm{y}_1,\ldots,\bm{y}_n) \right\},
\end{align*}
where $\bm{\theta}=(\xi,\bm{\beta}^\top)^\top$, $t_{1}(\bm{y}_1,\ldots,\bm{y}_n)=\sum_{i=1}^n\sum_{k\in N(i)}\sum_{j=1}^dy_{ij}y_{kj}$ and $t_l(\bm{y}_1,\ldots,\bm{y}_n)=\sum_{i=1}^n y_{il}$ for $l=2,\ldots,d+1$. This auto model has an intractable normalizing constant and score matching can be used to estimate the parameters.

The score matching objective function for $p(\bm{y}_1,\ldots,\bm{y}_n|\bm{\theta})$ based on the discussion in Section \ref{sect::sm manifold} is in the form of 
\[
d_{\rm SM}(\bm{\theta})=\frac{1}{2}\bm{\theta}^\top \bm{W}\bm{\theta}-\bm{\theta}^\top \bm{d}.
\]
Note that the projection matrix $\bm{P}$ for the product of $(d-1)$-spheres is given by $\bm{P}=\bm{I}_{nd\times nd}-\bm{\Gamma}$, with $\bm{I}_{nd\times nd}$ being the $nd\times nd$ identity matrix and $\bm{\Gamma}={\rm diag}\left( \bm{y_1}\bm{y_1}^\top,\ldots,\bm{y_n}\bm{y_n}^\top \right)$. For each sufficient statistic $t_l(y)$, create a vector-valued function $\bm{u}_l=\bm{u}_l(\bm{y}_1,\ldots,\bm{y}_n)=\nabla t_l(\bm{y}_1,\ldots,\bm{y}_n)$ by taking its Euclidean gradient. Then $\bm{W}$ is an $(d+1)\times (d+1)$ matrix with entries $w_{l_1l_2}=\frac{1}{n}{\rm E}\left(  \bm{u}_{l_1}^\top \bm{P}\bm{u}_{l_2} \right)$. The vector $\bm{d}$ has entries $d_l=-n^{-1}{\rm E}(\Delta_M t_l(\bm{y}_1,\ldots,\bm{y}_n))$, where the Laplace-Beltrami operator is given by
\begin{align*}
  \Delta_M \sum_{i=1}^ny_{ij}=-(d-1) \sum_{i=1}^ny_{ij},~~{\rm and}~~\Delta_M \sum_{i=1}^n\sum_{k\in N(i)}\sum_{j=1}^dy_{ij}y_{kj}=-2(d-1) \sum_{i=1}^n\sum_{k\in N(i)}\sum_{j=1}^dy_{ij}y_{kj}.
\end{align*}

In Section \ref{sect::sm dependent} we apply model (\ref{vMF_auto_model}) to a spatial geochemical dataset and construct a Wald test for investigating whether spatial dependence is present.

\section{Generalized Score Matching for Ordinal Data}\label{sect::sm discrete}

In this section, we develop generalized score matching for ordinal data. For the sake of simplicity, we only present generalized score matching for independent ordinal data, where Section \ref{subsection::sm discrete} and Section \ref{subsection::sm discrete multi} cover univariate and multivariate ordinal data, respectively. The extension to dependent data follows a similar idea to what is presented in Section \ref{subsection::sm discrete multi}.

To deal with discrete data, \cite{lyu2012interpretation} noticed that, in the continuous case, $\nabla_i \log q_i(\bm{y}_i)=\nabla_i q_i(\bm{y}_i)/q_i(\bm{y}_i)$ where the gradient is a linear operator. For discrete data, \cite{lyu2012interpretation} then proposed a generalization of score matching in which a general linear operator $\mathcal{L}_i$ replaces the gradient operator. Thus, the generalized score matching objective function $D_{\mathcal{L}_{\rm full}}(q_*,p_*)$ is given by
\begin{align}\label{original gsm loss}
  D_{\mathcal{L}_{\rm full}}(q_*,p_*)=&\frac{1}{n}\sum_{i=1}^n{\rm E} \left(\left\|\frac{\mathcal{L}_i q_i(\bm{y}_i)}{q_i(\bm{y}_i)}-\frac{\mathcal{L}_i  p_i(\bm{y}_i|\bm{\theta})}{ p_i(\bm{y}_i|\bm{\theta})}\right\|^2\right),
\end{align}
where $\|\cdot\|$ denotes the Euclidean norm. The particular linear operator considered in \cite{lyu2012interpretation} is called the marginalization operator $\mathcal{M}_i$ for $i\in\{1,\ldots,n\}$. This operator is defined to be $\mathcal{M}_iq_i(\bm{y}_i)\coloneqq(\mathcal{M}_i^{(1)}q_i(\bm{y}_i),\ldots,\mathcal{M}_i^{(d)}q_i(\bm{y}_i))^\top$ with $\mathcal{M}_i^{(j)}q_i(\bm{y}_i)\coloneqq\sum_{y_{ij}}q_i(\bm{y}_i)$ for $i\in\{1,\ldots,n\}$, $j=1,\ldots,d$ and the sum associated with operator $\mathcal{M}_i^{(j)}$ is taken over the support of $y_{ij}$. We offer two examples to explain the limitations of approaches in \cite{hyvarinen2007some} and \cite{lyu2012interpretation}.
\begin{exmp}
For univariate count data, the marginalization operator in \cite{lyu2012interpretation} will give $\mathcal{M}_i q_i (y_i)=\sum_{y_i}q_i(y_i)=1$ for $i\in\{1,\ldots,n\}$. Thus the generalized score matching objective function becomes
\[
  D_{\mathcal{L}_{\rm full}}(q_*,p_*)=\frac{1}{n}\sum_{i=1}^n{\rm E} \left(\left\|\frac{1}{q_i(y_i)}-\frac{1}{ p_i(y_i|\bm{\theta})}\right\|^2\right),
\]
which still involves the intractable normalizing constant. A similar problem occurs in the ratio matching approach due to \cite{hyvarinen2007some} in the case of univariate data.
\end{exmp}
\begin{exmp}
For multivariate count data, the marginalization operator in \cite{lyu2012interpretation} involves calculations of the sum over the support of data, which is an infinite sum. That means the marginalization operator leads to a computationally infeasible objective function. A parallel issue arises with the ratio matching approach due to \cite{hyvarinen2007some} as the method is developed for multivariate binary data.
\end{exmp}
To tackle these issues, we propose a novel linear operator in this section which is called the forward difference operator and our proposed method works well for either univariate or multivariate ordinal data.

\subsection{Generalized Score Matching for Univariate Ordinal Data}\label{subsection::sm discrete}
Score matching for continuous data is designed to compare the slopes of the logarithms of the densities. However, the proposed variations of score matching for discrete data fail to explore such relationships between the true density and the parametrized density. We consider a novel linear operator $\Delta_i$ which is defined as $\Delta_ip_i(y_i|\bm{\theta})\coloneqq p_i(y_i^+|\bm{\theta})-p_i(y_i|\bm{\theta})$ for $i\in\{1,\ldots,n\}$, where $y_i^+$ denotes the next possible value for $y_i$, i.e., $y_i^+=y_i+1$ when $y_i$ is count data. If the range of $y_i$ is bounded for $i\in\{1,\ldots,n\}$, let $p_i(y_i^+|\bm{\theta})=0$ when $y_i$ is located at the upper boundary. Similarly, let $y_i^-$ be the previous possible value for $y_i$, i.e., $y_i^-=y_i-1$ for count data, and $p_i(y_i^-|\bm{\theta})=0$ at the lower boundary of the data domain. Note that this linear operator gives a discrete analogue of the slope of $p_i(y_i|\bm{\theta})$ at the point $y_i$. After omitting the constant in $\Delta_ip_i(y_i|\bm{\theta})/p_i(y_i|\bm{\theta})$, we can see that the basic principle in our method is to force the ratio $p_i(y_i^+|\bm{\theta})/p_i(y_i|\bm{\theta})$ to be as close as possible to the corresponding ratio given by the data, i.e., $q_i(y_i^+)/q_i(y_i)$.

To avoid the zero denominator in the slopes, we consider the following transformation (\cite{hyvarinen2007some}) of the slopes:
$t(u)=1/(1+u)$. This choice of $t$ plays an important role both in our proofs and in the proofs of \cite{hyvarinen2007some} and \cite{lyu2012interpretation}.
Now, any probability that is zero and leads to a ratio that is infinite will give a value of $t(\infty)=0$ for this transformation.

Therefore, we propose that the model is estimated by minimizing the following objective function for $q_*$ and $p_*$:
\begin{align}\label{discrete overall population function}
  D_{\rm GSM}(q_*,p_*)=&\frac{1}{n} \sum_{i=1}^n {\rm E}  \left[ \left\{ t\left( \frac{p_i(y_i^+|\bm{\theta})}{p_i(y_i|\bm{\theta})} \right) - t\left( \frac{q_i(y_i^+)}{q_i(y_i)} \right) \right\}^2 + \left\{ t\left( \frac{p_i(y_i|\bm{\theta})}{p_i(y_i^-|\bm{\theta})}\right) - t\left( \frac{q_i(y_i)}{q_i(y_i^-)} \right) \right\}^2 \right]
\end{align}
The following theorem will show that (\ref{discrete overall population function}) is tractable. Its proof, as with all proofs in this article, is given in the Appendix.
\begin{theorem}\label{thm:: tractable obj funct for GSM}
  The overall population objective function (\ref{discrete overall population function}) can be decomposed as $D_{\rm GSM}(q_*,p_*)=g(q_*)+d_{\rm GSM}(\bm{\theta})$, where $g(q_*)$ is a constant depending on $q_*$ but not on $\bm{\theta}$ and 
  \begin{align*}
    d_{\rm GSM}(\bm{\theta})=&\frac{1}{n}  \sum_{i=1}^n {\rm E} \Bigg\{ t\left( \frac{p_i(y_i^+|\bm{\theta})}{p_i(y_i|\bm{\theta})} \right)^2+ t\left( \frac{p_i(y_i|\bm{\theta})}{p_i(y_i^-|\bm{\theta})}\right)^2-2t\left( \frac{p_i(y_i^+|\bm{\theta})}{p_i(y_i|\bm{\theta})} \right)\Bigg\}.
  \end{align*}
\end{theorem}

It is worth noting that Theorem \ref{thm:: tractable obj funct for GSM} indicates that $D_{\rm GSM}(q_*,p_*)$ has a tractable empirical version, which is given by
\begin{align}\label{discrete empirical function}
  \hat{d}_{\rm GSM}(\bm{\theta})=\frac{1}{n}\sum_{i=1}^n\rho_i^{\rm GSM}(y_i|\bm{\theta}),
\end{align}
where
\begin{align}\label{rho gsm}
  \rho_i^{\rm GSM}(y_i|\bm{\theta})=t\left( \frac{p_i(y_i^+|\bm{\theta})}{p_i(y_i|\bm{\theta})} \right)^2+ t\left( \frac{p_i(y_i|\bm{\theta})}{p_i(y_i^-|\bm{\theta})}\right)^2-2t\left( \frac{p_i(y_i^+|\bm{\theta})}{p_i(y_i|\bm{\theta})} \right).
\end{align}
The generalized score matching estimator for $\bm{\theta}$ is then defined as
\begin{align}\label{gsm est def}
  \hat{\bm{\theta}}=\argmin_{\bm{\theta}}\hat{d}_{\rm GSM}(\bm{\theta}).
\end{align}

The following result is analogous to the consistency theorem in \cite{hyvarinen2007some}. 
\begin{theorem}\label{thm:: GSM unique solution}
  Assume that the model $p_i(y_i|\bm{\theta})$ is correct, i.e. there exits a $\bm{\theta}_0$ such that $p_i(y_i|\bm{\theta}_0)=q_i(y_i)$ for $i\in\{1,\ldots,n\}$. Suppose also that the model is identifiable, i.e. for each $\bm{\theta}\neq\bm{\theta}_0$, there exists a set of $y_i$ of positive probability under $p_i(y_i|\bm{\theta}_0)$ such that $p_i(y_i|\bm{\theta})\neq p_i(y_i|\bm{\theta}_0)$. Then, $D_{\rm GSM}(q_*,p_*)=0$ if and only if $\bm{\theta}=\bm{\theta}_0$, where $D_{\rm GSM}$ is defined in (\ref{discrete overall population function}).
\end{theorem}

\subsection{Generalized Score Matching for Multivariate Ordinal Data}\label{subsection::sm discrete multi}
To extend our proposed generalized score matching to multivariate cases, we consider a linear operator $\widetilde{\Delta}_i=(\Delta_i^{(1)}, \ldots , \Delta_i^{(d)})^\top$ where $\Delta_i^{(j)} p_i(\bm{y}_i \vert \bm{\theta})\coloneqq   p_i(\bm{y}_i^{(j_{+})}|\bm{\theta})-p_i(\bm{y}_i|\bm{\theta})$ with $\bm{y}_i^{(j_{+})}=(y_{i1},\ldots,y_{ij}^+,\ldots,y_{id})^\top$ for $i\in\{1,\ldots,n\}$ and $j=1,\ldots,d$. If the range of $\bm{y}_i$ is bounded, let $p_i(\bm{y}_i^{(j_{+})}|\bm{\theta})=0$ when $\bm{y}_i^{(j_{+})}$ is located out of the sample space. Similarly, let $\bm{y}_i^{(j_{-})}=(y_{i1},\ldots,y_{ij}^-,\ldots,y_{id})^\top$ and $p_i(\bm{y}_i^{(j_{-})}|\bm{\theta})=0$ when $\bm{y}_i^{(j_{-})}$ is located out of the domain. After using the same transformation in the univariate case, the population objective function is given by
  \begin{align*}
    D_{\rm GSM}(q_*,p_*)=\frac{1}{n}\sum_{i=1}^n{\rm E} \Bigg(\sum_{j=1}^d \Bigg[\left\{ t\left( \frac{p_i(\bm{y}_i^{(j_{+})}|\bm{\theta})}{p_i(\bm{y}_i|\bm{\theta})} \right) - t\left( \frac{q_i(\bm{y}_i^{(j_{+})})}{q_i(\bm{y}_i)} \right) \right\}^2 + \left\{ t\left( \frac{p_i(\bm{y}_i|\bm{\theta})}{p_i(\bm{y}_i^{(j_{-})}|\bm{\theta})}\right) - t\left( \frac{q_i(\bm{y}_i)}{q_i(\bm{y}_i^{(j_{-})})} \right) \right\}^2 \Bigg]\Bigg).
  \end{align*}
Generalized score matching for multivariate ordinal data has analogous theoretical properties to those in the univariate ordinal case. Detailed discussion of these properties can be found in Section \ref{sup sec::gsm multi discrete} of the Supplementary Material.

\section{Theoretical Results for Score Matching Estimators}\label{Section_4}
Since the proposed generalized score matching estimators given by (\ref{sm est def}) and (\ref{gsm est def}) are M-estimators, standard asymptotic theory may be applied; see e.g. \cite{van2000asymptotic}. Our main goals in this section are (a) to provide a convenient, unified treatment for the ordinal and continuous cases and (b) to develop in greater detail than has been done previously a hypothesis testing framework based on score matching estimators.

\subsection{Conditions for Asymptotic Results}

Before discussing the limiting behaviors of our proposed score matching estimators for regression-type models, we state the following conditions for deriving asymptotics of our score matching estimator. For the sake of simplification, we present conditions for generalized score matching for multivariate discrete data. For univariate data, we can change the bold notation $\bm{y}_i$ to $y_i$. In the following, denote the parameter space for $\bm{\theta}$ by $\Theta$ and let $\mathcal{I}_n(\bm{\theta})=-{\rm E}\left( \frac{\partial^2 \hat{d}(\bm{\theta})}{\partial \bm{\theta} \partial \bm{\theta}^\top}\right)$ and $\mathcal{J}_n(\bm{\theta})=n{\rm Cov}\left(\frac{\partial \hat{d}(\bm{\theta})}{\partial \bm{\theta}}\right)$ in each case $\hat{d}(\cdot)=\hat{d}_{\rm SM}(\cdot)$ and $\hat{d}=\hat{d}_{\rm GSM}(\cdot)$ where $\hat{d}_{\rm SM}$ and $\hat{d}_{\rm GSM}$ are defined in (\ref{emprical sm cts}) and (\ref{discrete empirical function}), respectively. Additionally, we consider the corresponding cases $\rho_i=\rho_i^{\rm SM}$ and $\rho_i=\rho_i^{\rm GSM}$ where $\rho_i^{\rm SM}$ and $\rho_i^{\rm GSM}$ are given in (\ref{rho sm}) and (\ref{rho gsm}), respectively.

\begin{description}
\item[(C1)] There exists an open subset $\mathcal{B}$ of $\Theta$ that contains the true parameter point $\bm{\theta}_0$ such that for almost all $\bm{y}_i$, $\rho_i(\bm{y}_i|\bm{\theta})$ admits the first derivatives $\frac{\partial \rho_i(\bm{y}_i|\bm{\theta})}{\partial \bm{\theta}}$ and ${\rm E}( \rho_i(\bm{y}_i|\bm{\theta}))<\infty$ for all $\bm{\theta}\in\mathcal{B}$. Furthermore, for $i\in\{1,\ldots,n\}$,
\begin{align*}
\sup_{\bm{\theta}\in\mathcal{B}}\max_{k}\left| \frac{\partial \rho_i(\bm{y}_i|\bm{\theta})}{\partial \theta_{k}} \right| &\leq M_1(\bm{y}_i),
\end{align*}
where $m_1={\rm E}[|M_1(\bm{y}_i)|]<\infty$;

\item[(C2)] $\mathcal{I}_n(\bm{\theta}_0)\to\mathcal{I}(\bm{\theta}_0)$ and $\mathcal{J}_n(\bm{\theta}_0)\to\mathcal{J}(\bm{\theta}_0)$ as $n\to\infty$. We assume that $\mathcal{I}(\bm{\theta}_0)$ and $\mathcal{J}(\bm{\theta}_0)$ are positive definite;

\item[(C3)] There exists a $\delta>0$ such that $\widetilde{M}_i(\bm{\theta}_0)\coloneqq {\rm E}\left[\left\|\frac{\partial \rho_i(\bm{y}_i|\bm{\theta}_0)}{\partial \bm{\theta}}\right\|^{2+\delta}\right]$ satisfies $\lim_{n\to \infty}n^{-1}\sum_{i=1}^n \widetilde{M}_i(\bm{\theta}_0)=0$;

\item[(C4)] There exists an open subset $\mathcal{B}$ of $\Theta$ that contains the true parameter point $\bm{\theta}_0$ such that for almost all $\bm{y}_i$, $\rho_i(\bm{y}_i|\bm{\theta})$ admits all third derivatives $\frac{\partial^3 \rho_i(\bm{y}_i|\bm{\theta})}{\partial \theta_{j_1}\partial \theta_{j_2}\partial \theta_{j_3}}$ for all $\bm{\theta}\in\mathcal{B}$. Furthermore, for $i\in\{1,\ldots,n\}$, 
\begin{align*}
\sup_{\bm{\theta}\in\mathcal{B}} \max_{j_1,j_2,j_3}\left| \frac{\partial^3 \rho_i(\bm{y}_i|\bm{\theta})}{\partial \theta_{j_1}\partial \theta_{j_2}\partial \theta_{j_3}} \right| &\leq M_2(\bm{y}_i),
\end{align*}
where $m_2={\rm E}[|M_2(\bm{y}_i)|^{2+\delta}]<\infty$ for some $\delta>0$.

\end{description}

Condition (C1) comes from the differentiation lemma in \cite{klenke2013probability} which ensures the interchange of integration and differentiation in an open neighborhood around $\bm{\theta}_0$. Condition (C2) is a standard condition for establishing the convergence of the Fisher information matrix and the covariance of the score functions. Condition (C3) is a Lyapounov condition which, in conjunction with Condition (C4), is commonly used in asymptotics for MLEs under the INID (independent but not necessarily identically distributed) setting (\cite{lee1998analysis}).

\subsection{Asymptotic Normality}\label{sect:: thm property}

The derivation of a central limit theorem (CLT) for the score matching estimator $\hat{\bm{\theta}}$ relies on several key steps, including the Taylor series expansion; the CLT for the score vector; and the weak law of large numbers of the Hessian matrix. The asymptotics of the score function and the Hessian matrix are studied in Proposition \ref{prop::clt score function} in Section \ref{sect::tech details}. The remaining part of the proof, namely the proof of the CLT presented in Theorem \ref{thm::clt cts sm}, is very similar to that of Theorem 1 in \cite{zou2021network} which makes use of the $n^{1/2}$-consistency result in \cite{fan2001variable}. The details are omitted.

\begin{theorem}\label{thm::clt cts sm}
Suppose that Conditions (C1)-(C4) in Section \ref{sect:: thm property} hold. Then $\sqrt{n}(\hat{\bm{\theta}}-\bm{\theta}_0)\stackrel{d}{\longrightarrow} N\left(\bm{0}, \mathcal{K}(\bm{\theta}_0) \right)$, where $\hat{\bm{\theta}}$ is given by (\ref{sm est def}) or (\ref{gsm est def}) and $\mathcal{K}(\bm{\theta}_0)=\mathcal{I}^{-1}(\bm{\theta}_0)\mathcal{J}(\bm{\theta}_0)\mathcal{I}^{-1}(\bm{\theta}_0)$.
\end{theorem}

In practice, $\mathcal{K}(\bm{\theta}_0)$ is unknown but, under the conditions of the theorem, $\mathcal{K}(\bm{\theta}_0)$ may be consistently estimated by 
\begin{equation}
\hat{\mathcal{K}}_n(\hat{\bm{\theta}})=\hat{\mathcal{I}}_n^{-1}(\hat{\bm{\theta}})\hat{\mathcal{J}}_n(\hat{\bm{\theta}})\hat{\mathcal{I}}_n^{-1}(\hat{\bm{\theta}}), 
\label{curly_K_hat}
\end{equation}
where $\hat{\mathcal{I}}_n(\hat{\bm{\theta}})$ and $\hat{\mathcal{J}}_n(\hat{\bm{\theta}})$ are the sample analogues of $\mathcal{I}_n(\hat{\bm{\theta}})$ and $\mathcal{J}_n(\hat{\bm{\theta}})$, respectively.

%both $\mathcal{I}(\bm{\theta}_0)$ and $\mathcal{J}(\bm{\theta}_0)$ are unknown. To make the above theorem useful, one needs to find consistent estimators of these matrices. Using the fact that $\mathcal{I}_n(\bm{\theta}_0)\to \mathcal{I}(\bm{\theta}_0)$ and $\mathcal{J}_n(\bm{\theta}_0)\to \mathcal{J}(\bm{\theta}_0)$, we can show that the asymptotic covariance matrix $\mathcal{I}^{-1}(\bm{\theta}_0)+\mathcal{I}^{-1}(\bm{\theta}_0)\mathcal{J}(\bm{\theta}_0)\mathcal{I}^{-1}(\bm{\theta}_0)$ can be consistently estimated by $\mathcal{I}^{-1}_n(\hat{\bm{\theta}})+\mathcal{I}_n^{-1}(\hat{\bm{\theta}})\mathcal{J}_n(\hat{\bm{\theta}})\mathcal{I}_n^{-1}(\hat{\bm{\theta}})$. Additionally, it would be more convenient in practice to use  which are the sample analogues of $\mathcal{I}_n(\hat{\bm{\theta}})$ and $\mathcal{J}_n(\hat{\bm{\theta}})$, respectively.

\subsection{Hypothesis Testing}\label{sect::hypothesis testing}

Suppose $\bm{\theta}=(\bm{\theta}_1^\top ,  \bm{\theta}_2^\top)^\top$, where $\bm{\theta}_1$ is $\ell \times 1$ and $\bm{\theta}_2$ is $(p -\ell) \times 1$, with similar decompositions defined for $\bm{\theta}_0$ and $\hat{\bm{\theta}}$. The general form of the hypothesis test we wish to consider is
\begin{equation}
{\rm H}_0: \bm{\theta}_1=\bm{\theta}_{01}, \textrm{pre-specified}; \, \bm{\theta}_2 \, \textrm{unrestricted} \hskip 0.1truein \textrm{versus} \hskip 0.1truein {\rm H}_1: \bm{\theta}_1, \, \bm{\theta}_2 \, \textrm{both unrestricted}.
\label{gen_hyp}
\end{equation}
Inspired by commonly used tests within the maximum likelihood framework, we consider two tests for comparing the nested hypotheses in (\ref{gen_hyp}): a score-matching-based Wald test and a change-in-score-matching test. Then the score-matching-based Wald statistic for testing ${\rm H}_0$ in (\ref{gen_hyp}) is constructed as follows:
\begin{equation}
T_w=n(\hat{\bm{\theta}}_1 - \bm{\theta}_{01})^\top \hat{\mathcal{K}}_{n, 11}^{-1}(\hat{\bm{\theta}})(\hat{\bm{\theta}}_1 - \bm{\theta}_{01}) ,
\label{t_wald}
\end{equation}
where $\hat{\mathcal{K}}_{n,11}(\hat{\bm{\theta}})$ is the upper left $\ell \times \ell$ block of the matrix
\begin{equation}
\hat{\mathcal{K}}_n(\hat{\bm{\theta}})=\begin{pmatrix}
\hat{\mathcal{K}}_{n,11}(\hat{\bm{\theta}}) & \hat{\mathcal{K}}_{n,12}(\hat{\bm{\theta}}) \\
\hat{\mathcal{K}}_{n,21}(\hat{\bm{\theta}})  & \hat{\mathcal{K}}_{n,22}(\hat{\bm{\theta}}) 
\end{pmatrix}
\label{curly_K_2}
\end{equation}
defined in (\ref{curly_K_hat}). The following corollary follows from Theorem 3.
\begin{corollary}\label{coro::wald test INID}
Assume Conditions (C1)-(C4) in Section \ref{sect:: thm property} hold. Then, under the null hypothesis ${\rm H}_0$, we have $T_w\stackrel{d}{\longrightarrow}\chi^2_{\ell}$, as $n\to\infty$.
\end{corollary}
The change-in-score-matching test, a score-matching-based analogue of the likelihood ratio test, is now considered. Let $\widetilde{\bm{\theta}}=(\bm{\theta}_{01}^\top, \widetilde{\bm{\theta}}_2^\top)^\top$ denote the score matching estimator of $\bm{\theta}$ under the null hypothesis ${\rm H}_0$ in (\ref{gen_hyp}). The change-in-score-matching test statistic is 
\begin{equation}
T_c=-2n\left\{ \hat{d}(\widetilde{\bm{\theta}}) - \hat{d}(\hat{\bm{\theta}}) \right\}.
\label{change_test}
\end{equation}
Let $\mathcal{I}_{jk}(\bm{\theta}_0)$ and $\mathcal{J}_{jk}(\bm{\theta}_0)$, $j,k=1,2$ denote the four blocks of $\mathcal{I}(\bm{\theta}_0)$ and $\mathcal{J}(\bm{\theta}_0)$, respectively, corresponding to the block structure of (\ref{curly_K_2}), and define
\begin{equation}
\mathcal{A}(\bm{\theta}_0)=\mathcal{J}_{11}+\mathcal{I}_{12} \mathcal{I}_{22}^{-1} \mathcal{J}_{22}\mathcal{I}_{22}^{-1}\mathcal{I}_{21}
-\mathcal{I}_{12}\mathcal{I}_{22}^{-1}\mathcal{J}_{21}-\mathcal{J}_{12}\mathcal{I}_{22}^{-1}\mathcal{I}_{21},
\label{curly_A}
\end{equation}
where each term on the right of (\ref{curly_A}) is evaluated at $\bm{\theta}=\bm{\theta}_0$.  
%Recall that after rearrangement of $\bm{\theta}$, we have $\bm{\theta}=(\widetilde{\bm{\theta}}_1^\top,\widetilde{\bm{\theta}}_2^\top)^\top$ and 
%\[
%\mathcal{I}(\bm{\theta})=\begin{pmatrix} \mathcal{I}_{11}(\bm{\theta}) & \mathcal{I}_{12}(\bm{\theta}) \\
%\mathcal{I}_{21}(\bm{\theta}) & \mathcal{I}_{22}(\bm{\theta})   \end{pmatrix},
%\]
%where $\mathcal{I}_{ij}(\bm{\theta})$ is the convergence of its corresponding information matrix with respect to $\widetilde{\bm{\theta}}_i$ and $\widetilde{\bm{\theta}}_j$ for $i,j\in\{1,2\}$. Let $\widetilde{\mathcal{I}}_d(\bm{\theta})=\begin{pmatrix} \mathcal{I}_{11}^{-1}(\bm{\theta}) & \bm{0}_{(p-l)\times p}\\\bm{0}_{p\times(p-l)} &\bm{0}_{l\times l} \end{pmatrix}$, and denote $\mathcal{K}(\bm{\theta})=\mathcal{I}(\bm{\theta})+\mathcal{J}(\bm{\theta})$. Then the asymptotic distribution of $T_d$ is given below.
\begin{theorem}\label{thm::sm diff test}
Suppose that Conditions (C1)-(C4) in Section \ref{sect:: thm property} hold. The asymptotic distribution of the statistic $T_c$ in (\ref{change_test}) under $H_0$ in (\ref{gen_hyp}) is given by $\sum_{m=1}^\ell \lambda_m(\bm{\theta}_0)Z_m^2$ as $n \rightarrow \infty$, where $Z_1, \ldots , Z_\ell$ are independent $N(0,1)$ random variables and $\lambda_1(\bm{\theta}_0), \ldots , \lambda_{\ell}(\bm{\theta}_0)$ are the eigenvalues of the $\ell \times \ell$ matrix 
\begin{equation}
\mathcal{A}^{1/2}(\bm{\theta}_0) \left \{\mathcal{I}_{11}(\bm{\theta}_0) - \mathcal{I}_{12}(\bm{\theta}_0)\mathcal{I}_{22}(\bm{\theta}_0)^{-1}\mathcal{I}_{21}(\bm{\theta}_0)  \right \}^{-1} \mathcal{A}^{1/2}(\bm{\theta}_0),
\label{required_matrix}
\end{equation}
where $\mathcal{A}^{1/2}(\bm{\theta}_0)$ is the symmetric positive-definite square root of $\mathcal{A}(\bm{\theta}_0)$.
\end{theorem}

In practice, we can estimate $\lambda_1(\bm{\theta}_0), \ldots , \lambda_{\ell}(\bm{\theta}_0)$ by $\hat{\lambda}_1(\hat{\bm{\theta}}), \ldots , \hat{\lambda}_{\ell}(\hat{\bm{\theta}})$, the eigenvalues of the consistent estimator of the matrix (\ref{required_matrix}) obtained by replacing $\mathcal{I}_{jk}(\bm{\theta}_0)$ and $\mathcal{J}_{jk}(\bm{\theta}_0)$ in (\ref{curly_A}) and (\ref{required_matrix}) by their sample analogues $\hat{\mathcal{I}}_{n, jk}(\hat{\bm{\theta}})$ and $\hat{\mathcal{J}}_{n, jk}(\hat{\bm{\theta}})$, $j,k \in \{1,2\}$. A second point to note is that when $\ell=1$, (\ref{curly_A}) and (\ref{required_matrix}) are both real-valued.

In our later numerical work we compare the practical performance of the tests based on (\ref{t_wald}) and (\ref{change_test}).

\section{Numerical Study}\label{sect::total numerical study}
In this section we first analyze data on the number of Ph.D. biochemists' publications and then conduct Monte Carlo studies based on this dataset.

\subsection{Case Study: Doctoral Publication Data Analysis}\label{sect::case study}
\cite{long1990origins} studied the relationship between the number of Ph.D. student's publications and the gender (coded one for female), the marriage status (coded one if married), the number of children under age six (\emph{kid5}), the prestige of Ph.D. program (\emph{phd}) and the number of articles by mentor in last three years (\emph{mentor}). We focused only on the students with at least one publication and the data consists of 640 Ph.D. candidates. After selecting the 640 students with at least one publication, we subtracted one from the number of the publications for each sample. Then the average number of publications is 2.42. The number of publications exhibits strong over-dispersion (\cite{long2006regression}). We compared fits to the original data using the Conway-Maxwell-Poisson (CMP) regression model with approximate MLE and generalized score matching estimation. Both models were fit using a Windows desktop with an AMD Ryzen 9 5900X CPU running at 3.7 GHz and 32 GB RAM. The approximate MLE was obtained by using the function \code{glm.cmp} from the \textsf{R} package \texttt{COMPoissonReg} accompanying the paper by \cite{sellers2010flexible} and the generalized score matching estimator was performed using the function \code{optim} in the \textsf{R} package \texttt{stats} (\cite{R-Core-Team:2013wf}) with the Nelder-Mead algorithm (\cite{nelder1965simplex}).

The CMP regression model under independence is given by $p(y_1,\ldots,y_n|\bm{\theta})=\prod_{i=1}^n p_i(y_i|\bm{\theta})$, where $p_i(y_i|\bm{\theta})=\lambda_i^{y_i}/\{(y_i!)^\nu Z_i(\bm{\theta})\}$, $y_i\in\mathbb{N}_0$ with $\mathbb{N}_0$, the set of non-negative integers, $y_i!$ is $y_i$ factorial, $\bm{\theta}=(\bm{\beta}^\top,\nu)^\top$, $\nu\geq 0$ denotes the dispersion parameter and $\lambda_i=\exp (\bm{x}_i^\top\bm{\beta})$ is a generalization of the Poisson mean parameter for $i\in\{1,\ldots,n\}$ (though $\lambda_i$ is not itself the mean when $\nu\neq 1$) with $\bm{x}_i$ being the corresponding covariate vector for the $i$-th observation. The CMP regression model links together three common distributions as special cases: the geometric ($\nu=0$, and $\lambda_i<1$), Poisson ($\nu=1$), and Bernoulli ($\nu\to\infty$). Additionally, when $\nu\in[0,1)$, the CMP model describes over-dispersed data relative to a Poisson distribution with the same mean, while when $\nu>1$, the CMP model is appropriate for under-dispersed data (\cite{shmueli2005useful}).

The normalizing constant of the CMP regression model, $Z_i(\bm{\theta})=\sum_{s=0}^\infty \frac{\lambda_i^s}{(s!)^\nu}$, involves an infinite sum and is intractable. Therefore, within the maximum likelihood framework, several approximation approaches have been proposed, i.e., use of a truncation and an asymptotic approximation (\cite{shmueli2005useful}) of the normalization constant series. We refer to the resulting estimator as the approximate MLE. However, these approximations will become inaccurate under some situations. For example, the asymptotic approximation is accurate only if $\lambda_i>10^\nu$. Note that the function \code{glm.cmp} uses a hybrid method that includes the truncation and asymptotic approximations of the normalizing constant.

To avoid the explicit computation of this intractable normalizing constant, we conduct our novel generalized score matching and the empirical objective function of the CMP regression model is given by $\hat{d}_{\rm GSM}(\bm{\theta})=n^{-1}\sum_{i=1}^n \rho_i^{\rm GSM}(y_i|\bm{\theta})$, where 
\[
\rho_i^{\rm GSM}(y_i|\bm{\theta})\coloneqq\rho^{\rm GSM}(y_i|\bm{x}_i,\bm{\theta})=t\left(\frac{\lambda_i}{(y_i+1)^\nu}\right)^2+t\left(\frac{\lambda_i}{y_i^\nu}\right)^2-2t\left(\frac{\lambda_i}{(y_i+1)^\nu}\right).
\]

\begin{table}[htbp!]
\caption{Comparison of the approximate MLE and generalized score matching estimations for the CMP regression model with the doctoral publication dataset.\label{cmp::case study2}}
\centering
\scalebox{0.8}{
\begin{tabular}{ll | rrr | rrr}
  \hline
&&\multicolumn{3}{c|}{Approximate MLE}  & \multicolumn{3}{c}{Generalized score matching} \\
& Coefficient & Estimate & SE & $|z|$ & Estimate & SE & $|z|$ \\ 
   \hline
& intercept            &   -0.3345 &0.0712  & 4.6955 &-0.3141 &0.1022& 3.0736\\
& gender(Female).        &0.0097 &0.0594 & 0.1633 &-0.0893 & 0.0749 &  1.1931 \\
& marriage(Married)    & 0.0993& 0.0678 & 1.4654& 0.0445 & 0.0844 & 0.5268 \\
& kid5                  &-0.0726 &0.0339 & 2.1388   &  -0.0705 &0.0421 & 1.6747 \\
& phd                 & -0.0132 &0.0281 & 0.4687 & 0.0693 & 0.0394  & 1.7583  \\
& mentor                 & 0.1553 &0.0204 & 7.6252 & 0.0830 & 0.0347  & 2.3925  \\
\hline
&dispersion &   0.3698 &0.0402 & & 0.2564 &  0.0827 &\\
& run time (seconds) && 1.55   &&&0.70 &
\end{tabular}}
\end{table}

The estimated coefficients, standard errors (SEs), absolute z-statistics, dispersion and computer run times from the CMP regression model with approximate MLE and score matching estimation are given in Table \ref{cmp::case study2}. It is worth noting that the SEs of generalized score matching estimates are obtained based on the consistent estimator of the asymptotic variance, which is carefully discussed in Section \ref{sect:: thm property}.

For the standard CMP model, the interpretation of coefficients is opaque (\cite{Huang2017}), but we can ascertain the direction and statistical significance of the effect of each covariate from the fitted model. For example, Ph.D. candidates tend to have significantly more publications when their mentors have more publications in the last three years. First, we mention that the generalized score matching approach took only 0.7 seconds when fitting 7 parameters, which is much faster than the approximate MLE when fitting the same model. Second, the standard errors of the approximate MLE in Table \ref{cmp::case study2} are smaller than those obtained for generalized score matching. However, as seen in Table \ref{cmp coverage} in the next section, inference produced by the approximate MLE is not reliable.

\begin{figure}
\centering

\includegraphics[width=1\linewidth]{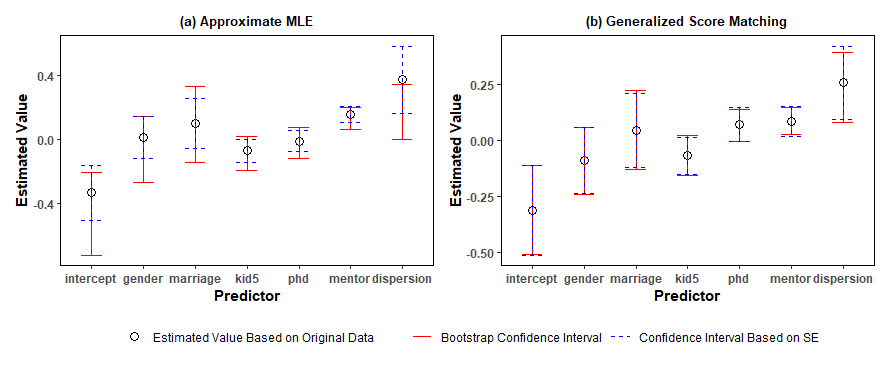}

\caption{Plots of $95\%$ parametric bootstrap confidence intervals and confidence intervals based on SE for fitted CMP regression models by (a) approximate MLE and (b) generalized score matching.}
\label{fig::case study ci}
\end{figure}

To study the standard errors of these two estimation methods, we first constructed confidence intervals based on the SE and the results are plotted in blue in Figure \ref{fig::case study ci}. The confidence interval of an estimated coefficient $\hat{\theta}_j$ based on its SE $s_j$ is given by $(\hat{\theta}_j-1.96s_j,\hat{\theta}_j+1.96s_j)$ for $j=1,\ldots,7$, where the estimated coefficients and SEs are reported in Table \ref{cmp::case study2}. We then consider the $95\%$ parametric bootstrap confidence intervals based on bootstrap percentile (\cite{Efron1994}). We generated 1000 bootstrap samples from the fitted CMP regression model based on generalized score matching and approximate MLE, respectively. For each bootstrap sample, the CMP parameters were re-estimated. Then 95\% confidence intervals for each unknown CMP parameters were constructed using the 1000 bootstrap quantiles for the corresponding parameter estimation. The results of the $95\%$ parametric bootstrap confidence intervals based on bootstrap percentile are given in Figure \ref{fig::case study ci} and plotted in red. It can be readily seen that, in Figure \ref{fig::case study ci}, the bootstrap confidence intervals and the confidence intervals based on SE overlap for generalized score matching. However, the length of the bootstrap confidence intervals is always larger than the one of the confidence intervals based on SE for approximate MLE. Figure \ref{fig::case study ci} indicates that the approximate MLE method underestimates the variance of the parameters and yields biased estimates.

For model diagnostics, a uniform quantile plot is given in Figure \ref{fig::case study} in Section \ref{supp subsect::simulation cmp} of the Supplementary Material. This plot shows reasonable closeness to uniformity, which indicates that the fitted CMP regression model using generalized score matching is appropriate. To examine the prediction accuracy of the fitted models using our generalized score matching estimators and approximate MLE, we randomly split the whole data set into training and test data sets where the training set contained 448 samples and the test set contained 192 observations. The predicted number of publications was obtained by taking the mean of simulated data from the CMP model with estimated parameters. It is worth noting that simulating data from a CMP by using the approximation of the normalizing constant is not proper in our numerical study as the approximation may be inaccurate. We decided to use the rejection sampling algorithm by \cite{chanialidis2018efficient}, which is an exact method to generate CMP data. The test MSE for the model fitted by generalized score matching is 2.64 and the test MSE for the model using the approximate MLE estimation is 3.90. The comparison of the test MSEs indicate better performance of our proposed generalized score matching method.

\subsection{Simulation Studies}\label{sect::numerical study}
We conducted a numerical study to evaluate the performance of the generalized score matching estimator for ordinal data based on the dataset discussed in Section \ref{sect::case study}. We are particularly interested in examining the bias, standard deviation and the root mean squared error of the score matching estimator. In addition, we compare our generalized score matching approach with the approximate MLE approach. 

All simulations were conducted via $1000$ replicates. For the purpose of assessing the performance of parameter estimators, we denote $\hat{\bm{\theta}}^{(k)}$ as the vector estimation of $\bm{\theta}$ in the $k$-th replicate. For each component of $\bm{\theta}$, which is $\theta_j$, the averaged bias of $\hat{\theta}_j^{(k)}$, $k\in\{1,\ldots,1000\}$, is $\textrm{BIAS}=1000^{-1}\sum_{k}(\hat{\theta}_j^{(k)}-\theta_j)$, and the standard deviation of $\hat{\theta}_j^{(k)}$ is $\textrm{SD}=\Big\{1000^{-1}\sum_{k_1}(\hat{\theta}_j^{(k_1)}-1000^{-1}\sum_{k_2}\hat{\theta}_j^{(k_2)})^2\Big\}^{1/2}$. Therefore, the root mean squared error is $\textrm{RMSE}=\sqrt{\textrm{SD}^2+\textrm{BIAS}^2}$.

For $i\in\{1,\ldots,n\}$, we considered the $6\times 1$ covariate vector $\bm{x}_i$ which was randomly sampled with replacement from the original publication dataset in Section \ref{sect::case study}, and their corresponding regression parameters were the generalized score matching estimator of the fitted CMP model in Section \ref{sect::case study}, that is $\bm{\beta}_0=(\beta_1,\beta_2,\beta_3,\beta_4,\beta_5,\beta_6)^\top=(-0.3141,-0.0893,0.0445,$ $-0.0705,0.0693,0.0830)^\top$. The true dispersion parameter was set to be $\nu_0=0.2564$ and the covariate matrix is fixed across the replications.

\begin{figure}
\centering

\includegraphics[width=1\linewidth]{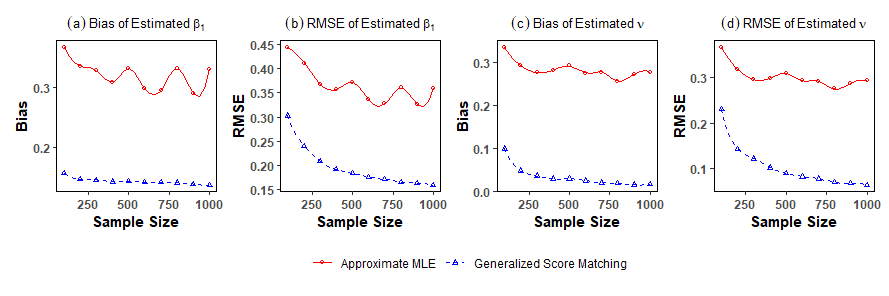}

\caption{Plots of (a) BIAS, (b) RMSE of estimated $\beta_1$ for fitted CMP regression models by approximate MLE and generalized score matching and (c) BIAS, (d) RMSE of estimated $\nu$ for fitted CMP regression models by approximate MLE and generalized score matching.}
\label{fig::numerical study plot}
\end{figure}

For the CMP regression model, Figure \ref{fig::numerical study plot} shows trends of BIAS and RMSE of the generalized score matching estimation and approximate MLE. To save space, we only present the results for $\beta_1$ and $\nu$. The detailed results for BIAS, SD and RMSE can be found in Table \ref{normal:sim4} in Section \ref{supp subsect::simulation cmp} of the Supplementary Material and Table \ref{normal:sim4} yields similar findings. In Figure \ref{fig::numerical study plot}, the sample size $n$ varied in $\{100,200,\ldots,1000\}$ and we find that the BIAS and RMSE for generalized score matching generally become smaller for all parameter estimates as $n$ becomes larger. The above findings support our theoretical results that the generalized score matching estimator for ordinal data is consistent. Moreover, we notice that the generalized score matching method is much more accurate when the approximate MLE is biased. Figure \ref{fig::numerical study plot} indicates that the approximation of the normalizing constant is inaccurate and thus the bias of the approximate MLE does not always decrease when the sample size $n$ becomes larger. Additionally, after comparing the trends of the bias and the RMSE, it is seen that the approximate MLE typically has a small SD and a large bias.

\begin{table}[H]
\caption{Comparison of the generalized score matching estimations and approximate MLEs of the parameters ($\beta_1=-0.3141$, $\beta_2=-0.0893$, $\beta_3=0.0445$, $\beta_4=-0.0705$, $\beta_5=0.0693$, $\beta_6=-1.3610$, $\nu=0.2564$) for the CMP regression model. One measure is considered: the empirical coverages of a $95\%$ confidence interval constructed by generalized score matching estimations and approximate MLEs, respectively.\label{cmp coverage}}
\centering
\scalebox{0.8}{
\begin{tabular}{lcclrrrr}
  \hline
&Estimation&$n$&Measure& $\beta_1$ & $\beta_2$ & $\beta_3$ & $\beta_4$  \\ 
  \hline
&\multirow{3}{*}{Score matching}&$n=200$&Coverage& 0.9280 &0.9460 &0.9260 &0.9460 \\
  &&$n=500$ &Coverage & 0.9420 &0.9440& 0.9360& 0.9540   \\
  &&$n=1000$ &Coverage& 0.9580 &0.9480 &0.9540& 0.9580 \\
  \hline
&\multirow{3}{*}{Approximate MLE}&$n=200$&Coverage& 0.7600& 0.7900& 0.8060& 0.8820 \\
  &&$n=500$ &Coverage & 0.4820& 0.6160& 0.7240 &0.7460   \\
  &&$n=1000$ &Coverage& 0.3980 &0.4120& 0.6280 &0.6860  \\
  \hline
&Estimation&$n$&Measure& $\beta_5$ & $\beta_6$ & $\nu$  \\ 
  \hline
  &\multirow{3}{*}{Score matching}&$n=200$&Coverage& 0.9580 &0.9580&0.9600 \\
  &&$n=500$  &Coverage & 0.9420 &0.9460 &0.9460   \\
  &&$n=1000$ &Coverage & 0.9520 &0.9520 &0.9580\\
   \hline
&\multirow{3}{*}{Approximate MLE}&$n=200$&Coverage& 0.8500 &0.8260 &0.8200 \\
  &&$n=500$  &Coverage &0.7140 &0.7100 &0.4320  \\
  &&$n=1000$ &Coverage &0.6260 &0.5380&0.2220\\
   \hline
\end{tabular}
}
\end{table}

We investigated the empirical coverages of a $95\%$ confidence interval constructed by generalized score matching estimation and approximate MLE, respectively. The results are given in Table \ref{cmp coverage}. The sample size $n$ we considered in the following varied in $\{200,500,1000\}$. These results indicate that the generalized score matching estimator is asymptotically normal while inference based on the approximate MLE is not reliable. Due to the inaccurate approximation of the normalizing constant, even though the SD decreases when the sample size gets larger, the bias and RMSE remain relatively large and, consequently, the empirical coverages based on the approximate MLE are rather poor, even with a larger sample size.

The above findings support our theoretical results that the generalized score matching estimation is consistent and asymptotically normal. While exact maximum likelihood estimation is intractable and the approximate MLE is biased and performs poorly in some situations, in contrast generalized score matching produces consistent estimation and reliable inference.

We next assess the finite sample performance of the score-matching-based Wald test and the change-in-score-matching test via evaluating the empirical size with the significance levels ranging from 0.01 to 0.30 and the empirical power with the significance level 0.05. The empirical size and power are the percentages of rejections under ${\rm H}_0$ and ${\rm H}_1$, respectively via the hypothesis test ${\rm H}_0: \widetilde{\bm{\beta}}=(\beta_2,\beta_3,\beta_4,\beta_5)^\top=\bm{0}~~{\rm versus}~~{\rm H}_1:\widetilde{\bm{\beta}}\neq\bm{0}$, with $1000$ realizations. Note that this hypothesis test can be represented by the one discussed in Section \ref{sect:: thm property} by rearranging $\bm{\theta}$. The empirical size is the percentage of rejections under the setting of $(\beta_1,\beta_2,\beta_3,\beta_4,\beta_5,\beta_6)=(-0.3141,0,0,0,0,0.0830)$ and $\nu_0=0.2564$. The empirical power is the percentage of rejections under the setting of $(\beta_1,\beta_2,\beta_3,\beta_4,\beta_5,\beta_6)=(-0.3141,-0.0893\iota,0.0445\iota,-0.0705\iota,0.0693\iota,0.0830)$ and $\nu_0=0.2564$, where the signal strength $\iota>0$.

\begin{figure}
\centering
\includegraphics[width=1\linewidth]{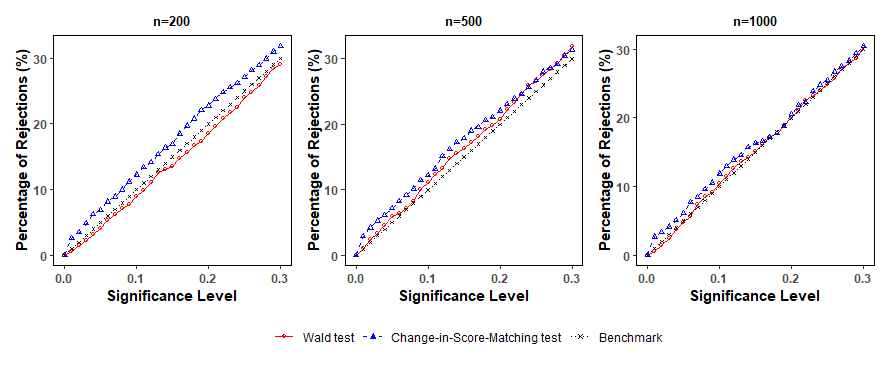}
\caption{The empirical sizes of the score-matching-based Wald test and the change-in-score-matching test for the significance levels ranging from 0.01 to 0.30 under the setting of the CMP regression model. The benchmark represents the ideal case when the percentage of rejections from 1000 replications is equal to the significance level.}
\label{fig::simulation cmp test compare size}
\end{figure}

\begin{figure}
\centering
\includegraphics[width=1\linewidth]{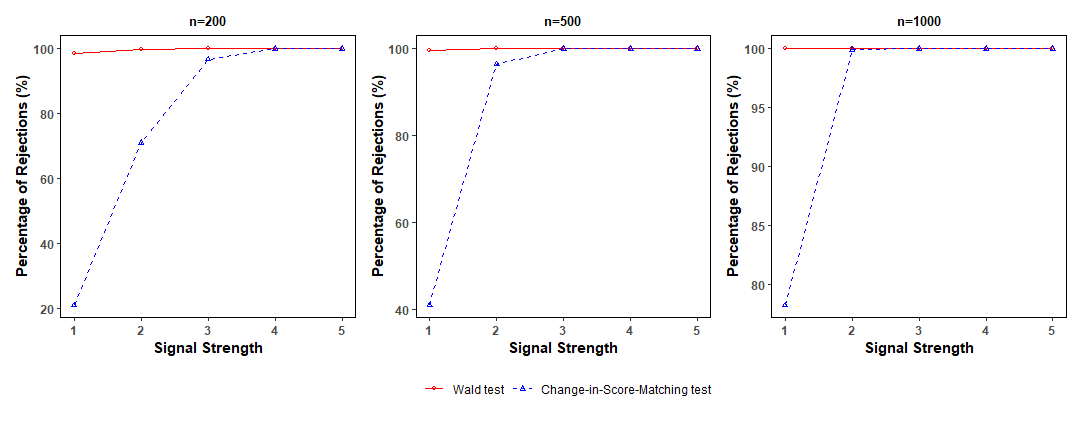}
\caption{The empirical power of the score-matching-based Wald test and the change-in-score-matching test at a nominal level of 0.05. The signal strengths $\iota=1,2,3,4$ and 5 which correspond to the settings $(\beta_1,\beta_2,\beta_3,\beta_4,\beta_5,\beta_6)=(-0.3141,-0.0893\iota,0.0445\iota,-0.0705\iota,$ $0.0693\iota,0.0830)$, respectively.}
\label{fig::simulation cmp test compare power}
\end{figure}

Figure \ref{fig::simulation cmp test compare size} shows that the empirical sizes of the score-matching-based Wald test and the change-in-score-matching test are almost identical to the predetermined significance levels at $n=1000$. Additionally, Figure \ref{fig::simulation cmp test compare size} indicates that the change-in-score-matching test is oversized (anticonservative) when $n$ is not large enough, but the score-matching-based Wald test enables us to control the size reasonably well, especially at the significance level 0.05. Figure \ref{fig::simulation cmp test compare power} shows the empirical powers of these two tests tend to 100\% when the sample size $n$ or the signal strength $\iota$ gets larger. However, we find that the change-in-score-matching test is not powerful when the signal strength $\iota$ is small. These findings indicate that these two tests perform well when $n$ is large and the score-matching-based Wald test performs much better when $n$ is small.

While our numerical studies exclusively focus on the univariate setting when dealing with ordinal data, we anticipate that our proposed method will demonstrate strong performance in multivariate scenarios. This expectation is supported by the findings of numerical studies conducted in \cite{meng2022concrete,matsubara2023generalized}. The numerical study on score matching for continuous data yields similar findings. The detailed results can be found in Section \ref{supp subsect::simulation truncated gaussian} of the Supplementary Material.

\section{Score Matching for Dependent Data}\label{sect::sm dependent}
In this section, we return to the vMF auto model for spherical data introduced in Section \ref{Section_2.3}. It is worth noting that due to the complex dependence structure of auto models, general asymptotic results are not currently available. However, it is the case that the gradient score matching objective function is unbiased at the correct model. In this section we propose a score-matching-based Wald statistic to test for the presence of spatial dependence and we apply our model and methodology to study a geochemical dataset from Spain.

To study the spatial dependence of the observations, we consider the null and alternative hypotheses given by ${\rm H}_0:\xi=0~~{\rm versus}~~{\rm H}_1:\xi\neq 0$. For testing ${\rm H}_0$, we choose the score-matching-based Wald test discussed in Section \ref{sect::hypothesis testing}, and the test statistic is $T_w=n\hat{\xi}^2/\hat{\mathcal{K}}_{n,11}(\hat{\bm{\theta}})$, where $\hat{\mathcal{K}}_{n,11}(\hat{\bm{\theta}})$ is the upper left $1\times1$ block of the matrix $\hat{\mathcal{K}}_n(\hat{\bm{\theta}})$, $\hat{\mathcal{K}}_n(\hat{\bm{\theta}})=n \left( \hat{\bm{W}}\hat{\bm{\theta}}-\hat{\bm{d}} \right)\left( \hat{\bm{W}}\hat{\bm{\theta}}-\hat{\bm{d}} \right)^\top$, $\hat{\bm{W}}$ and $\hat{\bm{d}}$ are the sample version of $\bm{W}$ and $\bm{d}$ given in Section \ref{Section_2.3}, respectively. The asymptotic distribution of $T_w$ is given in the following theorem.

\begin{theorem}\label{thm::wald test}
  Suppose that ${\rm E}\left( |y_{ij_1}y_{ij_2}y_{ij_3}y_{ij_4}|^{2+\eta} \right)<\infty$ for some $\eta>0$, $i\in\{1,\ldots,n\}$ and $j_1,j_2,j_3,j_4\in\{1,\ldots,d\}$. Further assume that there exist positive definite matrices $\bm{W}$ and $\mathcal{K}(\bm{\theta}_0)=\lim_{n\to\infty}\mathcal{K}_n(\bm{\theta}_0)$ where $\mathcal{K}_n(\bm{\theta}_0)=n{\rm E}\left[  \left( \hat{\bm{W}}\bm{\theta}_0-\hat{\bm{d}} \right)\left( \hat{\bm{W}}\bm{\theta}_0-\hat{\bm{d}} \right)^\top \right]$. Then, under the null hypothesis ${\rm H}_0$, we have, as $n \rightarrow \infty$, $T_w\stackrel{d}{\longrightarrow}\chi^2_1$.
\end{theorem}

To illustrate the advantage of our vMF proposed auto models, we studied a geochemical data on agricultural and grazing land soil. This dataset was collected in the GEMAS project (\cite{reimann2014chemistry,reimann2014chemistry2}) and consisted of measurement of chemical elements on land soil in Europe. We focused only on the data in Spain and the main chemical elements were Al, Ca, Fe, K and Si. The major elements were reported in weight percent by collecting 202 samples of agricultural soil. These samples were collected from geographically dispersed sites and these site locations satisfied the grid-based setting. The neighbourhood structure we considered for the grid-based setting is the first order neighbour structure, that is, for the site $i$ which is located in the central of Spain, it has eight neighbours. After the square root transformation for the compositional data, the observations were located on the positive orthant of the sphere $\mathcal{S}^5$ with high concentration and none of the observations were located at the boundary. Therefore, our proposed auto model is suitable for this dataset. The estimated values are $\hat{\bm{\beta}}=(2.8792,2.3916,1.9828,1.5974,6.5620,1.6320)^\top$ and $\hat{\xi}=4.8566$. The score-matching-based Wald test gives $T_w=8.5409$ and the corresponding p-value is 0.0034, which indicates that we should reject the null hypothesis. We conclude that there exists the spatial dependence between these observations.

We then compared our auto model with an independent model, that is, we directly used the vMF distribution to fit the data. Note that based on our auto model, the conditional density of each observation $\bm{y}_i$ is also a vMF distribution with the estimated concentration parameter $\hat{\kappa}_i=\|\hat{\bm{\beta}}+\hat{\xi}\sum_{k\in N(i)}\bm{y}_k\|$ and the estimated mean direction
\[
\hat{\bm{\beta}}_i=\frac{\hat{\bm{\beta}}+\hat{\xi}\sum_{k\in N(i)}\bm{y}_k }{\|\hat{\bm{\beta}}+\hat{\xi}\sum_{k\in N(i)}\bm{y}_k\|}.
\]

\begin{figure}
\centering

\includegraphics[width=1\linewidth]{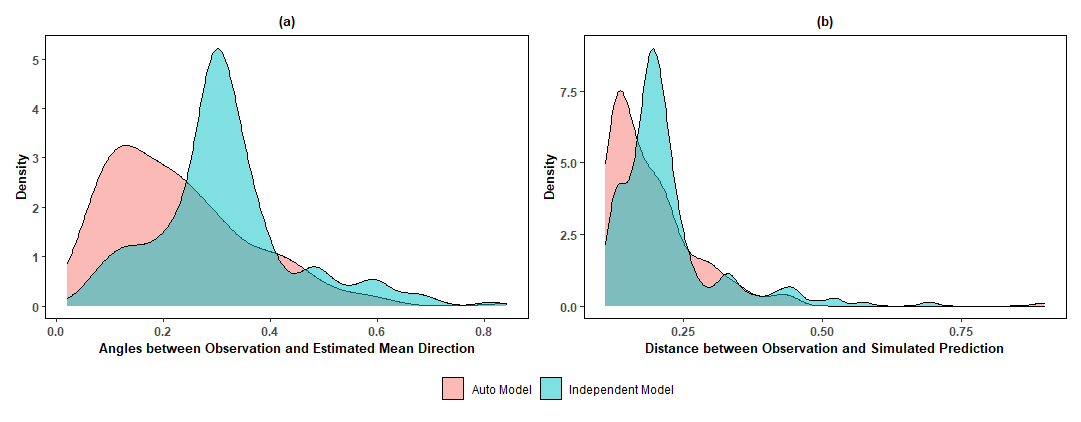}
  
\caption{Density plots for auto model and independent model with measures produced by (a) angles (radian) between the observations and the estimated mean direction; (b) distance between the observations and the simulated predictions}
\label{fig:auto compare}
\end{figure}

We first compared angles between observations and the estimated mean direction for the fitted auto model and independent model. The first plot in Figure \ref{fig:auto compare} indicated that most of the angles between the observations and the estimated mean direction produced by the conditional density of our auto model were close to 0.2 radian while most of the observations did not lie on the same direction along with the estimated mean direction for the independence model. For each observation, we then generated 1000 samples independently by the conditional density of the fitted auto model and the independent model, respectively. The second plot in Figure \ref{fig:auto compare} reported the average Euclidean distance between the observations and the simulated predictions for the auto model and independent model, respectively. This measure can be viewed as a training MSE and the second plot Figure \ref{fig:auto compare} also indicates a better fit for our auto model.

\section{Conclusion}

In this article, we extend score matching  beyond continuous IID models. Specifically, we propose a novel generalized score matching approach for ordinal data. The proposed generalized score matching approach goes beyond previous research in that it can be applied to univariate and multivariate ordinal data. The simulations and real data analysis support our theoretical results. Furthermore, our proposed generalized score matching technique has the potential to make significant contributions to the fields of Bayesian statistics and deep learning based on the subsequent studies in \cite{meng2022concrete,matsubara2023generalized}. By deriving the consistency and asymptotic normality of the proposed estimators under the independence assumption, we establish the theoretical foundation for score-matching-based inference for general models. Additionally, we propose a novel auto model for spherical data and develop a score-matching-based Wald test to test the spatial independence. This illustration also shows that the extension of score matching beyond the IID case advances the fields of both statistical estimation and statistical modeling.

\section{Technical Details}\label{sect::tech details}

\subsection{Auxiliary Results and Their Proofs}\label{sect::tech proposition}

Our first result in this section shows that the expectation of the score vector is equal to $\bm{0}$ when $\bm{\theta}=\bm{\theta}_0$. Note that $\bm{\theta}_0$ is the population value of $\bm{\theta}$ assuming that the parametric model is correct. 

\begin{proposition}\label{thm:: sm score function expectation}
Suppose that Condition (C1) in Section \ref{sect:: thm property} holds. Then
\[
{\rm E}\left(\frac{\partial \hat{d}(\bm{\theta}_0)}{\partial \bm{\theta}}\Bigg|_{\bm{\theta}=\bm{\theta}_0}\right)=\bm{0},
\]
in each case $\hat{d}(\cdot)=\hat{d}_{\rm SM}(\cdot)$ and $\hat{d}=\hat{d}_{\rm GSM}(\cdot)$ where $\hat{d}_{\rm SM}$ and $\hat{d}_{\rm GSM}$ are defined in (\ref{emprical sm cts}) and (\ref{discrete empirical function}), respectively, $\bm{\theta}_0$ denotes the population parameter vector and the expectation is taken with respect to the population distribution.
\end{proposition}

\begin{proof}[Proof of Proposition \ref{thm:: sm score function expectation}]
This proof consists of two parts. First, we consider the case when $\hat{d}(\cdot)=\hat{d}_{\rm SM}(\cdot)$ and then we derive results for $\hat{d}(\cdot)=\hat{d}_{\rm GSM}(\cdot)$.

\emph{Part I}: When $\hat{d}(\cdot)=\hat{d}_{\rm SM}(\cdot)$, $\hat{d}_{\rm SM}(\bm{\theta})=\frac{1}{n}\sum_{i=1}^n \rho_i^{\rm SM}(\bm{y}_i|\bm{\theta})$, we have
\begin{align}\label{exp eq 1}
{\rm E}\left(\frac{\partial \hat{d}_{\rm SM}(\bm{\theta})}{\partial \bm{\theta}} \right)&=\frac{1}{n}\sum_{i=1}^n {\rm E}\left(\frac{\partial \rho_i^{\rm SM}(\bm{y}_i|\bm{\theta})}{\partial \bm{\theta}} \right) =\frac{1}{n}\sum_{i=1}^n\frac{\partial}{\partial\bm{\theta}}{\rm E}\left[ \rho_i^{\rm SM}(\bm{y}_i|\bm{\theta}) \right].
\end{align}
The last equality holds based on the interchange of integration and differentiation under Condition (C1). After interchanging the differentiation and integral sign, (\ref{exp eq 1}) then implies that
\begin{align*}
{\rm E}\left(\frac{\partial \hat{d}_{\rm SM}(\bm{\theta})}{\partial \bm{\theta}} \right)&=\frac{1}{n}\sum_{i=1}^n \frac{\partial}{\partial\bm{\theta}} {\rm E}\left( \|\nabla_{i}\log q_i(\bm{y}_i)-\nabla_{i}\log p_i(\bm{y}_i|\bm{\theta})\|^2 \right)\\
&= \frac{1}{n}\sum_{i=1}^n  {\rm E}\left(  \frac{\partial}{\partial\bm{\theta}} \|\nabla_{i}\log q_i(\bm{y}_i)-\nabla_{i}\log p_i(\bm{y}_i|\bm{\theta})\|^2 \right)\\
&= -\frac{1}{n}\sum_{i=1}^n  {\rm E}\left( 2\nabla_{i}\left(\log q_i(\bm{y}_i)-\log p_i(\bm{y}_i|\bm{\theta})\right)  \frac{\partial}{\partial\bm{\theta}} \nabla_{i}\log p_i(\bm{y}_i|\bm{\theta}) \right).
\end{align*}
It can be readily seen that ${\rm E}\left(\frac{\partial \hat{d}_{\rm SM}(\bm{\theta}_0)}{\partial \bm{\theta}}\right)=\bm{0}$ since $\log q_i(\bm{y}_i)=\log p_i(\bm{y}_i|\bm{\theta}_0)$ for $i\in\{1,\ldots,n\}$. This completes the first part of proof.

\emph{Part II}: When $\hat{d}(\cdot)=\hat{d}_{\rm GSM}(\cdot)$, under Condition (C1), we have
\begin{align*}
{\rm E}\left( \frac{\partial \hat{d}_{\rm GSM}(\bm{\theta})}{\partial \bm{\theta}} \right) =&\frac{1}{n}\sum_{i=1}^n \frac{\partial}{\partial \bm{\theta}}{\rm E}\left[ \rho_i^{\rm GSM}(y_i|\bm{\theta})\right]\\
=& \frac{1}{n}\sum_{i=1}^n \sum_{y\in\mathcal{D}} q_i(y_i) \frac{\partial}{\partial \bm{\theta}}  \Bigg\{\Bigg[ t\left( \frac{p_i(y_i^+|\bm{\theta})}{p_i(y_i|\bm{\theta})} \right) - t\left( \frac{q_i(y_i^+)}{q_i(y_i)} \right) \Bigg]^2 + \Bigg[ t\left( \frac{p_i(y_i|\bm{\theta})}{p_i(y_i^-|\bm{\theta})}\right) - t\left( \frac{q_i(y_i)}{q_i(y_i^-)} \right) \Bigg]^2 \Bigg\}\\
=&\frac{1}{n}\sum_{i=1}^n \sum_{y\in\mathcal{D}} q_i(y_i)  \Bigg\{ 2\Bigg[ t\left( \frac{p_i(y_i^+|\bm{\theta})}{p_i(y_i|\bm{\theta})} \right) - t\left( \frac{q_i(y_i^+)}{q_i(y_i)} \right) \Bigg] \frac{\partial}{\partial \bm{\theta}} \Bigg[ t\left( \frac{p_i(y_i^+|\bm{\theta})}{p_i(y_i|\bm{\theta})} \right) - t\left( \frac{q_i(y_i^+)}{q_i(y_i)} \right) \Bigg]\\
& + 2\Bigg[ t\left( \frac{p_i(y_i|\bm{\theta})}{p_i(y_i^-|\bm{\theta})}\right) - t\left( \frac{q_i(y_i)}{q_i(y_i^-)} \right) \Bigg]\frac{\partial}{\partial \bm{\theta}} \Bigg[ t\left( \frac{p_i(y_i|\bm{\theta})}{p_i(y_i^-|\bm{\theta})}\right) - t\left( \frac{q_i(y_i)}{q_i(y_i^-)} \right) \Bigg] \Bigg\}.
\end{align*}
The proof is completed by noting that
\[
{\rm E}\left(\frac{\partial \hat{d}_{\rm GSM}(\bm{\theta}_0)}{\partial \bm{\theta}}\right)=\bm{0}.
\]
\end{proof}

The above result gives 
\begin{align*}
\mathcal{J}_n(\bm{\theta}_0)=n{\rm Cov}\left(\frac{\partial \hat{d}(\bm{\theta}_0)}{\partial \bm{\theta}}\right)=\frac{1}{n}\sum_{i=1}^n {\rm E}\left\{\frac{\partial \rho_i(\bm{y}_i|\bm{\theta}_0)}{\partial \bm{\theta}}\frac{\partial \rho_i(\bm{y}_i|\bm{\theta}_0)}{\partial \bm{\theta}^\top}\right\}.
\end{align*}

\begin{proposition}\label{prop::clt score function}
Suppose that Conditions (C1)-(C4) in Section \ref{sect:: thm property} hold. Then
\[
\sqrt{n}\frac{\partial \hat{d}(\bm{\theta}_0)}{\partial \bm{\theta}}\stackrel{d}{\longrightarrow}N\left(\bm{0},\mathcal{J}(\bm{\theta}_0)\right),
\]
and
\[
-\frac{\partial^2 \hat{d}(\bm{\theta}_0)}{\partial \bm{\theta} \partial \bm{\theta}^\top}\stackrel{p}{\longrightarrow}\mathcal{I}(\bm{\theta}_0);
\]
in each case $\hat{d}(\cdot)=\hat{d}_{\rm SM}(\cdot)$ and $\hat{d}=\hat{d}_{\rm GSM}(\cdot)$ where $\hat{d}_{\rm SM}$ and $\hat{d}_{\rm GSM}$ are defined in (\ref{emprical sm cts}) and (\ref{discrete empirical function}), respectively, $\mathcal{I}(\bm{\theta}_0)$ and $\mathcal{J}(\bm{\theta}_0)$ which are defined in Condition (C2) in Section \ref{sect:: thm property} are the limits of $\mathcal{I}_n(\bm{\theta}_0)$ and $\mathcal{J}_n(\bm{\theta}_0)$, respectively.
\end{proposition}

\begin{proof}[Proof of Proposition \ref{prop::clt score function}]
We use the Lyapounov theorem for triangular arrays to derive our results. For the proof of the first part of Proposition \ref{prop::clt score function}, we first consider $\hat{d}=\hat{d}_{\rm SM}$ and define a triangular array
\[
\bm{z}_{ni}=\frac{\partial \rho_i^{\rm SM}(\bm{y}_i|\bm{\theta}_0)}{\partial \bm{\theta}},
\]
for $i\in\{1,\ldots,n\}$. Note that we have ${\rm E}(\bm{z}_{ni})=\bm{0}$ and $\bm{z}_{ni}$ is actually a function of $\bm{y}_i$. Thus  $\bm{z}_{n1},\ldots,\bm{z}_{nn}$ are INID random variables with mean $\bm{0}$. Note that the Condition (C2) gives
\[
\frac{1}{n}{\rm Cov}\left(\sum_{i=1}^n\bm{z}_{ni}\right)=\mathcal{J}_n(\bm{\theta}_0)\to\mathcal{J}(\bm{\theta}_0),
\]
by Lyapounov theorem and the Lyapounov condition (C3), we get
\[
n^{\frac{1}{2}}\frac{\partial \hat{d}_{\rm SM}(\bm{\theta}_0)}{\partial \bm{\theta}}=n^{-\frac{1}{2}}\sum_{i=1}^n\bm{z}_{ni}\stackrel{d}{\longrightarrow}N\left(\bm{0},\mathcal{J}(\bm{\theta}_0)\right).
\]

For the proof of the second part of Proposition \ref{prop::clt score function}, we slightly abuse the notion of $\bm{z}_{ni}$ and use this to define a new triangle array
\[
\bm{z}_{ni}={\rm vec}\left(\frac{\partial^2 \rho_i^{\rm SM}(\bm{y}_i|\bm{\theta}_0)}{\partial \bm{\theta} \partial \bm{\theta}^\top}\right).
\] 
Under Condition (C4), one can be readily seen that
\begin{align*}
\sup_n \max_i {\rm E}\|\bm{z}_{ni}\|^2= \sup_n\max_i {\rm E}\left(\left\|\frac{\partial^2 \rho_i^{\rm SM}(\bm{y}_i|\bm{\theta}_0)}{\partial \bm{\theta} \partial \bm{\theta}^\top}\right\|_F^2 \right)
<\infty,
\end{align*}
thus by the weak law of large numbers and Condition (C2), we obtain
\[
-\frac{\partial^2 \hat{d}_{\rm SM}(\bm{\theta}_0)}{\partial \bm{\theta} \partial \bm{\theta}^\top}\stackrel{p}{\longrightarrow}\mathcal{I}(\bm{\theta}_0),
\]
which completes the proof for $\hat{d}=\hat{d}_{\rm SM}$. After employing similar methods to those used in the proof for $\hat{d}=\hat{d}_{\rm SM}$, we could finish the proof for $\hat{d}=\hat{d}_{\rm GSM}$.
\end{proof}

\begin{proposition}\label{prop::quadratic clt}
For INID random vectors $\bm{y}_i\in\mathbb{R}^d$ with $i\in\{1,\ldots,n\}$, mean $\bm{\mu}_i$ and given sets $\{N(i):i=1\ldots,n\}$, define $\bm{Q}_n=\left( \sum_{i=1}^n \bm{y}_i^\top ,\sum_{i=1}^n\sum_{k\in N(i)}\bm{y}_i^\top\bm{y}_k\right)^\top $, let $\bm{\mu}_{\bm{Q}_n}$ be the mean of $\bm{Q}_n$ and $\bm{\Sigma}_{\bm{Q}_n}$ be the covariance matrix of $\bm{Q}_n$, then $\bm{\mu}_{\bm{Q}_n}=\left( \sum_{i=1}^n \bm{\mu}_i^\top , \sum_{i=1}^n\sum_{k\in N(i)}\bm{\mu}_i^\top\bm{\mu}_k \right)^\top$ and
\[
\bm{\Sigma}_{\bm{Q}_n}=\begin{pmatrix} \bm{\Sigma}_{\bm{Q}_n}^{(11)} & \bm{\Sigma}_{\bm{Q}_n}^{(12)}\\
\bm{\Sigma}_{\bm{Q}_n}^{(21)} & \Sigma_{\bm{Q}_n}^{(22)}
 \end{pmatrix},
\]
where 
\begin{align*}
\bm{\Sigma}_{\bm{Q}_n}^{(11)}&=\sum_{i=1}^n \left( \bm{\Sigma}_i-\bm{\mu}_i\bm{\mu}_i^\top \right) ,\\
\bm{\Sigma}_{\bm{Q}_n}^{(12)}&=2\sum_{i=1}^n\sum\limits_{ \substack{k\in N(i)\\k< i}} \left( \bm{\Sigma}_i\bm{\mu}_k-\bm{\mu}_i\bm{\mu}_i^\top \bm{\mu}_k \right),\\
\bm{\Sigma}_{\bm{Q}_n}^{(21)}&=\bm{\Sigma}_{\bm{Q}_n}^{(12)\top},\\
\Sigma_{\bm{Q}_n}^{(22)}&=4\sum_{i=1}^n \left[ \sum\limits_{ \substack{k\in N(i)\\k< i}}{\rm tr}(\bm{\Sigma}_k\bm{\Sigma}_i) + \sum\limits_{ \substack{k_1,k_2\in N(i)\\k_1\neq k_2< i}} \bm{\mu}_{k_1}^\top \bm{\Sigma}_i \bm{\mu}_{k_2} - \left( \sum\limits_{ \substack{k\in N(i)\\k< i}} \bm{\mu}_i^\top\bm{\mu}_k \right) \left( \sum\limits_{ \substack{k\in N(i)\\k< i}} \bm{\mu}_k\bm{\mu}_i^\top \right) \right],
\end{align*}
with $\bm{\Sigma}_i={\rm E}(\bm{y}_i\bm{y}_i^\top)$. Moreover, $n^{-1/2-\varepsilon}(\bm{Q}_n-\bm{\mu}_{\bm{Q}_n})\stackrel{L_2}{\longrightarrow}\bm{0}$ for any $\varepsilon>0$. Further assume that there exists a positive definite matrix $\bm{\Sigma}\in\mathbb{R}^{(d+1)\times (d+1)}$ such that $n^{-1}\bm{\Sigma}_{\bm{Q}_n}\to \bm{\Sigma}$. Then
\[
n^{-1/2}(\bm{Q}_n-\bm{\mu}_{\bm{Q}_n})\stackrel{d}{\longrightarrow} N(\bm{0},\bm{\Sigma}).
\]
\end{proposition}

\begin{proof}[Proof of Proposition \ref{prop::quadratic clt}]
First note that 
\[
\bm{Q}_n-\bm{\mu}_{\bm{Q}_n}=\sum_{i=1}^n \bm{z}_{i,n},
\]
where 
\[
\bm{z}_{i,n}=\begin{pmatrix}  \bm{y}_i-\bm{\mu}_i \\ 2\sum\limits_{ \substack{k\in N(i)\\k< i}} \bm{y}_k^\top\bm{y}_i -2\sum\limits_{ \substack{k\in N(i)\\k< i}} \bm{\mu}_k^\top\bm{\mu}_i \end{pmatrix}.
\]
Consider the $\sigma$-field $\mathcal{F}_{0,n}=\{\emptyset,\Omega\}$, $\mathcal{F}_{i,n}=\sigma(\bm{y}_j:j\in N(1)\cup\ldots\cup N(i) \cap \{1,\ldots,i\} )$, $1\leq i\leq n$. By construction $\mathcal{F}_{i-1,n}\subseteq \mathcal{F}_{i,n}$, $\bm{z}_{i,n}$ is $\mathcal{F}_{i,n}$-measurable, and ${\rm E}(\bm{z}_{i,n}|\mathcal{F}_{i-1,n})=0$. Thus $\{ \bm{z}_{i,n},\mathcal{F}_{i,n},1\leq i\leq n, n\geq 1 \}$ forms a martingale difference array. Therefore $\bm{\Sigma}_{\bm{Q}_n}=\sum_{i=1}^n {\rm E} (\bm{z}_{i,n}\bm{z}_{i,n}^\top)$. The expression of $\bm{\Sigma}_{\bm{Q}_n}$ is given by
\[
\bm{\Sigma}_{\bm{Q}_n}=\begin{pmatrix} \bm{\Sigma}_{\bm{Q}_n}^{(11)} & \bm{\Sigma}_{\bm{Q}_n}^{(12)}\\
\bm{\Sigma}_{\bm{Q}_n}^{(21)} & \Sigma_{\bm{Q}_n}^{(22)}
 \end{pmatrix},
\]
where 
\begin{align*}
\bm{\Sigma}_{\bm{Q}_n}^{(11)}&=\sum_{i=1}^n \left( \bm{\Sigma}_i-\bm{\mu}_i\bm{\mu}_i^\top \right) ,\\
\bm{\Sigma}_{\bm{Q}_n}^{(12)}&=2\sum_{i=1}^n\sum\limits_{ \substack{k\in N(i)\\k< i}} \left( \bm{\Sigma}_i\bm{\mu}_k-\bm{\mu}_i\bm{\mu}_i^\top \bm{\mu}_k \right),\\
\bm{\Sigma}_{\bm{Q}_n}^{(21)}&=\bm{\Sigma}_{\bm{Q}_n}^{(12)\top},\\
\Sigma_{\bm{Q}_n}^{(22)}&=4\sum_{i=1}^n \left[ \sum\limits_{ \substack{k\in N(i)\\k< i}}{\rm tr}(\bm{\Sigma}_k\bm{\Sigma}_i) + \sum\limits_{ \substack{k_1,k_2\in N(i)\\k_1\neq k_2< i}} \bm{\mu}_{k_1}^\top \bm{\Sigma}_i \bm{\mu}_{k_2} - \left( \sum\limits_{ \substack{k\in N(i)\\k< i}} \bm{\mu}_i^\top\bm{\mu}_k \right) \left( \sum\limits_{ \substack{k\in N(i)\\k< i}} \bm{\mu}_k\bm{\mu}_i^\top \right) \right],
\end{align*}
with $\bm{\Sigma}_i={\rm E}(\bm{y}_i\bm{y}_i^\top)$. Clearly, we have ${\rm Cov}\left[ n^{-1/2-\varepsilon}(\bm{Q}_n-\bm{\mu}_{\bm{Q}_n}) \right]\to\bm{0}$ for any $\varepsilon>0$. This implies 
\[
n^{-1/2-\varepsilon}(\bm{Q}_n-\bm{\mu}_{\bm{Q}_n})\stackrel{L_2}{\longrightarrow}\bm{0}.
\]

Let $\widetilde{\bm{z}}_{i,n}=\bm{\Sigma}_{\bm{Q}_n}^{-1/2}\bm{z}_{i,n}$, then $\{ \widetilde{\bm{z}}_{i,n},\mathcal{F}_{i,n},1\leq i\leq n, n\geq 1 \}$ also forms a martingale difference array. We now prove that
\[
\bm{\Sigma}_{\bm{Q}_n}^{-1/2}\left( \bm{Q}_n-\bm{\mu}_{Q_n} \right)=\sum_{i=1}^n \widetilde{\bm{z}}_{i,n} \stackrel{d}{\longrightarrow} N(\bm{0},I_d),
\]
by showing that $\widetilde{\bm{z}}_{i,n}$ satisfies the remaining conditions of Lemma \ref{vector mda clt} in the Supplementary Material.

Take $0<\delta<\eta/2$, we note that under the maintained moment conditions on $\bm{y}_i$, there exist some finite constant $K_j$ and $M_{jk}$ for $j,k\in\{1,\ldots,d\}$ such that $\sigma_{ij}={\rm E}(y_{ij}-\mu_{ij})^2\leq K_j$ and ${\rm E}(|y_{ij}|^s){\rm E}(|y_{kj}|^s)\leq M_{jk}$ for $s\leq 2+\delta$. Note that there always exist some constant $\widetilde{M}_j$ such that $\max_{1\leq k\leq n} M_{jk} \leq \widetilde{M}_j$. Let $\widetilde{N}_i=|N(i)\backslash \{i,\ldots,n\}|$ and $q=2+\delta$, we have
\begin{align*}
\|\bm{z}_{i,n}\|^q&\leq 2^q \|\bm{y}_i-\mu_i \|^q+2^q \Bigg| 2\sum\limits_{ \substack{k\in N(i)\\k< i}} \bm{y}_k^\top\bm{y}_i -2\sum\limits_{ \substack{k\in N(i)\\k< i}} \bm{\mu}_k^\top\bm{\mu}_i \Bigg|^q\\
&\leq (2d)^q \sum_{j=1}^d |y_{ij}-\mu_{ij} |^q+2^{3q} (\widetilde{N}_id)^q \sum\limits_{ \substack{k\in N(i)\\k< i}} \sum_{j=1}^d\left( |y_{kj}|^q|y_{ij}|^q+|\mu_{kj}|^q|\mu_{ij}|^q \right).
\end{align*}
Consequently,
\begin{align*}
&\sum_{i=1}^n{\rm E}\left\{{\rm E}\left[ \| \bm{z}_{i,n} \|^q|\mathcal{F}_{i-1,n} \right] \right\}\\
 \leq &(2d)^q \sum_{i=1}^n\sum_{j=1}^d {\rm E}( |y_{ij}-\mu_{ij} |^q)+2^{3q} d^q \sum_{i=1}^n \widetilde{N}_i^q \sum\limits_{ \substack{k\in N(i)\\k< i}}\sum_{j=1}^d\left( {\rm E}( |y_{kj}|^q) {\rm E}(|y_{ij}|^q)+|\mu_{kj}|^q|\mu_{ij}|^q \right)\\
 \leq & n\left[ (2d)^q \sum_{j=1}^d K_j + 2^{3q+1} d^q(\sup_{n\geq 1}\max_{1\leq i \leq n}\widetilde{N}_i)^{q+1}\sum_{j=1}^d \widetilde{M}_j\right].
\end{align*}
Therefore, we can conclude that 
\begin{align*}
\sum_{i=1}^n{\rm E}\left\{{\rm E}\left[ \| \widetilde{\bm{z}}_{i,n} \|^q|\mathcal{F}_{i-1,n} \right] \right\}\leq & n^{1+\delta/2} \|\bm{\Sigma}_{\bm{Q}_n}^{-1/2}\|^{2+\delta} \frac{1}{n^{1+\delta/2}} \sum_{i=1}^n{\rm E}\left\{{\rm E}\left[ \| \bm{z}_{i,n} \|^q|\mathcal{F}_{i-1,n} \right] \right\}\\
\leq & \left(n \|\bm{\Sigma}_{\bm{Q}_n}^{-1}\|\right)^{1+\delta/2} \frac{1}{n^{\delta/2}} \left[ (2d)^q \sum_{j=1}^d K_j + 2^{3q+1} d^q(\sup_{n\geq 1}\max_{1\leq i \leq n}\widetilde{N}_i)^{q+1}\sum_{j=1}^d\widetilde{M}_j\right].
\end{align*}
Note that based on the condition on $\bm{\Sigma}_{\bm{Q}_n}$, we can always find a sufficient large $n$ such that $n\|\bm{\Sigma}_{\bm{Q}_n}^{-1}\|\leq c$ for some constant $c>0$. Thus the r.h.s of the last inequality goes to zero as $n\to\infty$, which shows that Condition (\ref{vector mda clt condition1 suff condition}) holds.

Note that 
\[
{\rm E}\left[ \bm{z}_{i,n}\bm{z}_{i,n}^\top |\mathcal{F}_{i-1,n} \right]=\begin{pmatrix} \bm{\Sigma}_{\mathcal{F}}^{(11)} & \bm{\Sigma}_{\mathcal{F}}^{(12)} \\ \bm{\Sigma}_{\mathcal{F}}^{(21)} & \Sigma_{\mathcal{F}}^{(22)} \end{pmatrix},
\]
where
\begin{align*}
\bm{\Sigma}_{\mathcal{F}}^{(11)}&=  \bm{\Sigma}_i-\bm{\mu}_i\bm{\mu}_i^\top  ,\\
\bm{\Sigma}_{\mathcal{F}}^{(12)}&=2\sum\limits_{ \substack{k\in N(i)\\k< i}} \left( \bm{\Sigma}_i\bm{y}_k-2\bm{\mu}_i\bm{\mu}_i^\top \bm{y}_k +\bm{\mu}_i\bm{\mu}_i^\top \bm{\mu}_k \right),\\
\bm{\Sigma}_{\mathcal{F}}^{(21)}&=\bm{\Sigma}_{\mathcal{F}}^{(12)\top},\\
\Sigma_{\mathcal{F}}^{(22)}&=4 \left[ \sum\limits_{ \substack{k_1,k_2\in N(i)\\k_1,k_2< i}} \bm{y}_{k_1}^\top \bm{\Sigma}_i \bm{y}_{k_2} -2\left( \sum\limits_{ \substack{k\in N(i)\\k< i}} \bm{\mu}_i^\top\bm{y}_k \right) \left( \sum\limits_{ \substack{k\in N(i)\\k< i}} \bm{\mu}_k\bm{\mu}_i^\top \right)+ \left( \sum\limits_{ \substack{k\in N(i)\\k< i}} \bm{\mu}_i^\top\bm{\mu}_k \right) \left( \sum\limits_{ \substack{k\in N(i)\\k< i}} \bm{\mu}_k\bm{\mu}_i^\top \right) \right].
\end{align*}
Recalling that $\bm{\Sigma}_{\bm{Q}_n}=\sum_{i=1}^n {\rm E} (\bm{z}_{i,n}\bm{z}_{i,n}^\top)$ and utilizing the expression for ${\rm E} (\bm{z}_{i,n}\bm{z}_{i,n}^\top)$ yields
\begin{align*}
\sum_{i=1}^n {\rm E}\left[ \widetilde{\bm{z}}_{i,n}\widetilde{\bm{z}}_{i,n}^\top |\mathcal{F}_{i-1,n} \right]-\bm{I}_d&=\bm{\Sigma}_{\bm{Q}_n}^{-1/2}\sum_{i=1}^n \left\{  {\rm E}\left[ \bm{z}_{i,n}\bm{z}_{i,n}^\top |\mathcal{F}_{i-1,n} \right] -  {\rm E} (\bm{z}_{i,n}\bm{z}_{i,n}^\top)\right\}  \bm{\Sigma}_{\bm{Q}_n}^{-1/2}\\
&= n^{1/2}\bm{\Sigma}_{\bm{Q}_n}^{-1/2} \begin{pmatrix} \bm{\Sigma}_D^{(11)} &\bm{\Sigma}_D^{(12)}\\ \bm{\Sigma}_D^{(21)} & \bm{\Sigma}_D^{(22)}  \end{pmatrix} n^{1/2} \bm{\Sigma}_{\bm{Q}_n}^{-1/2},
\end{align*}
where 
\begin{align*}
\bm{\Sigma}_{D}^{(11)}=&\bm{0}_{d\times d} ,\\
\bm{\Sigma}_{D}^{(12)}=&2n^{-1}\sum_{i=1}^n\sum\limits_{ \substack{k\in N(i)\\k< i}} \left( \bm{\Sigma}_i(\bm{y}_k-\bm{\mu}_k)-2\bm{\mu}_i\bm{\mu}_i^\top (\bm{y}_k-\bm{\mu}_k) \right),\\
\bm{\Sigma}_{D}^{(21)}=&\bm{\Sigma}_{D}^{(12)\top},\\
\Sigma_{D}^{(22)}=&4n^{-1}\sum_{i=1}^n \Bigg[ \sum\limits_{ \substack{k\in N(i)\\k< i}} {\rm tr}\left[(\bm{y}_k\bm{y}_k^\top-\bm{\Sigma}_k)\bm{\Sigma}_i\right] + \sum\limits_{ \substack{k_1,k_2\in N(i)\\k_1\neq k_2< i}}( \bm{y}_{k_1}- \bm{\mu}_{k_1})^\top \bm{\Sigma}_i ( \bm{y}_{k_2}- \bm{\mu}_{k_2})\\
&- 2\Bigg( \sum\limits_{ \substack{k\in N(i)\\k< i}} \bm{\mu}_i^\top(\bm{y}_k-\bm{\mu}_k) \Bigg) \Bigg( \sum\limits_{ \substack{k\in N(i)\\k< i}} \bm{\mu}_k\bm{\mu}_i^\top \Bigg) \Bigg].
\end{align*}
Clearly, $\bm{\Sigma}_{D}^{(12)} \stackrel{p}{\longrightarrow} \bm{0}$ by the weak law of large numbers. Utilizing the Cramer–Wold device and the weak law of large numbers for martingale different arrays in \cite{davidson1994stochastic} it follows that
\begin{align*}
&n^{-1}\sum_{i=1}^n  \sum\limits_{ \substack{k_1,k_2\in N(i)\\k_1\neq k_2< i}}( \bm{y}_{k_1}- \bm{\mu}_{k_1})^\top \bm{\Sigma}_i ( \bm{y}_{k_2}- \bm{\mu}_{k_2})  \stackrel{p}{\longrightarrow} 0,\\
&n^{-1}\sum_{i=1}^n  \left( \sum\limits_{ \substack{k\in N(i)\\k< i}} \bm{\mu}_i^\top(\bm{y}_k-\bm{\mu}_k) \right) \left( \sum\limits_{ \substack{k\in N(i)\\k< i}} \bm{\mu}_k\bm{\mu}_i^\top \right)  \stackrel{p}{\longrightarrow} 0,\\
&n^{-1}\sum_{i=1}^n \sum\limits_{ \substack{k\in N(i)\\k< i}} {\rm tr}\left[(\bm{y}_k\bm{y}_k^\top-\bm{\Sigma}_k)\bm{\Sigma}_i\right] \stackrel{p}{\longrightarrow} 0.
\end{align*}
The above results, together with the condition that $n\|\bm{\Sigma}_{\bm{Q}_n}^{-1}\|\leq c$, imply that Condition (\ref{vector mda clt condition2}) holds. This completes the entire proof.

\end{proof}

\subsection{Proofs of Main Results}
\begin{proof}[Proof of Theorem \ref{thm:: tractable obj funct for GSM}]
Recall that the transformation function $t(f_1(y)/f_2(y))$ will give a zero value when $f_2(y)=0$ and $t(f_1(y)/f_2(y))=1$ when $f_1(y)=0$. In other words, the transformation $t(f_1(y)/f_2(y))$ will return a constant when either $f_1(y)$ or $f_2(y)$ equals zero. Therefore, for the sake of simplicity, we assume that $q_i(y_i)$ and $p_i(y_i|\bm{\theta})$ are non-zero for $y\in\mathcal{D}$ where $\mathcal{D}$ is the support of the distribution. Note that we have $D_{\rm GSM}(q_*,p_*)=\sum_{i=1}^nD_{\rm GSM}(q_i,p_i)$, where
\begin{align*}
D_{\rm GSM}(q_i,p_i)=&\sum_{y\in\mathcal{D}}q_i(y_i) \Bigg\{ t\left( \frac{p_i(y_i^+|\bm{\theta})}{p_i(y_i|\bm{\theta})} \right)^2 + t\left( \frac{p_i(y_i|\bm{\theta})}{p_i(y_i^-|\bm{\theta})}\right)^2 \Bigg\}-2\sum_{y\in\mathcal{D}}q_i(y_i) \Bigg\{ t\left( \frac{p_i(y_i^+|\bm{\theta})}{p_i(y_i|\bm{\theta})} \right) t\left( \frac{q_i(y_i^+)}{q_i(y_i)} \right) \\
&  + t\left( \frac{p_i(y_i|\bm{\theta})}{p_i(y_i^-|\bm{\theta})}\right) t\left( \frac{q_i(y_i)}{q_i(y_i^-)} \right)  \Bigg\}+C_i,
\end{align*}
where $C_i$ does not depend on $\bm{\theta}$. We first consider the second term, which can be manipulated as follows:
\begin{align*}
&\sum_{y\in\mathcal{D}}q_i(y_i) \Bigg\{ t\left( \frac{p_i(y_i^+|\bm{\theta})}{p_i(y_i|\bm{\theta})} \right) t\left( \frac{q_i(y_i^+)}{q_i(y_i)} \right)+ t\left( \frac{p_i(y_i|\bm{\theta})}{p_i(y_i^-|\bm{\theta})}\right) t\left( \frac{q_i(y_i)}{q_i(y_i^-)} \right)  \Bigg\}\nonumber\\
=&\sum_{y\in\mathcal{D}}q_i(y_i) \Bigg\{ t\left( \frac{p_i(y_i^+|\bm{\theta})}{p_i(y_i|\bm{\theta})} \right)  \frac{q_i(y_i)}{q_i(y_i^+)+q_i(y_i)} +t\left( \frac{p_i(y_i|\bm{\theta})}{p_i(y_i^-|\bm{\theta})}\right) \frac{q_i(y_i^-)}{q_i(y_i^-) +q_i(y_i)}  \Bigg\}\\
=&\sum_{y\in\mathcal{D}}q_i(y_i) t\left( \frac{p_i(y_i^+|\bm{\theta})}{p_i(y_i|\bm{\theta})} \right)  \frac{q_i(y_i)}{q_i(y_i^+)+q_i(y_i)} + \sum_{y\in\mathcal{D}}q_i(y_i^+) t\left( \frac{p_i(y_i^+|\bm{\theta})}{p_i(y_i|\bm{\theta})}\right) \frac{q_i(y_i)}{q_i(y_i^+) +q_i(y_i)}\\
=& \sum_{y\in\mathcal{D}}q_i(y_i) t\left( \frac{p_i(y_i^+|\bm{\theta})}{p_i(y_i|\bm{\theta})} \right).
\end{align*}

Therefore, we can conclude that
\begin{align*}
D_{\rm GSM}(q_i,p_i)=\sum_{y\in\mathcal{D}}q_i(y_i)\Bigg\{ t\left( \frac{p_i(y_i^+|\bm{\theta})}{p_i(y_i|\bm{\theta})} \right)^2+ t\left( \frac{p_i(y_i|\bm{\theta})}{p_i(y_i^-|\bm{\theta})}\right)^2-2t\left( \frac{p_i(y_i^+|\bm{\theta})}{p_i(y_i|\bm{\theta})} \right) \Bigg\}+C_i,
\end{align*}
which completes the entire proof.

\end{proof}

\begin{proof}[Proof of Theorem \ref{thm:: GSM unique solution}]
Based on the analysis of the transformation function $t(\cdot)$ in the proof of Theorem \ref{thm:: tractable obj funct for GSM}, we assume that $q_i(y_i)$ and $p_i(y_i|\bm{\theta})$ are non-zero for $y\in\mathcal{D}$. The hypothesis $D_{\rm GSM}(q_*,p_*)=0$, in conjunction with the assumption that $p(y|\bm{x}_i,\bm{\theta}_0)=q_i(y_i)>0$, implies that all the slopes must be equal for the model and the observed data. Thus, we have
\begin{align}
\frac{p_i(y_i^+|\bm{\theta})}{p_i(y_i|\bm{\theta})}&=\frac{q_i(y_i^+)}{q_i(y_i)},\label{forward}\\
\frac{p_i(y_i|\bm{\theta})}{p_i(y_i^-|\bm{\theta})}&=\frac{q_i(y_i)}{q_i(y_i^-)} ,\label{backward}
\end{align} 
for all $y$ and $i\in\{1,\ldots,n\}$. It is worth noting that the relationships (\ref{forward}) and (\ref{backward}) are equivalent. Without loss of generality, we only consider the first case (\ref{forward}) with $y^+\in\mathcal{D}$. We then obtain
\[
\frac{q_i(y_i) }{p_i(y_i|\bm{\theta})}=\frac{q_i(y_i^+)}{p_i(y_i^+|\bm{\theta})}.
\]
Applying this identity on $y_i^+$, we get
\[
\frac{q_i(y_i^+)}{p_i(y_i^+|\bm{\theta})}=\frac{q(y^{++}|\bm{x}_i)}{p(y^{++}|\bm{x}_i,\bm{\theta})},
\]
for $y^{++}$ being the one after the next possible value and $y^{++}\in\mathcal{D}$. It can be readily seen that we can recursively apply this identity.

Now, fix any points $y_i^0$ for $i\in\{1,\ldots,n\}$. Without loss of generality, we assume that $y_i^0\geq y_i$. By using the recursion above, we have
\[
\frac{q_i(y_i) }{p_i(y_i|\bm{\theta})}=\frac{q_i(y_i^+)}{p_i(y_i^+|\bm{\theta})}=\cdots=\frac{q_i(y_i^0) }{p_i(y_i^0|\bm{\theta})}=c_i,
\]
where $c_i$ is a constant which does not depend on $y_i$. Therefore, we can conclude that $q_i(y_i)=c_i p_i(y_i|\bm{\theta})$ for any $y_i$.

On the other hand, both $p_i$ and $q_i$ are normalized probability distributions. Thus, we must have $c_i=1$. This proves that if $D_{\rm GSM}(q_*,p_*)=0$, then $q_i(y_i)=p_i(y_i|\bm{\theta})$ for any $y_i$ and $i\in\{1,\ldots,n\}$. Using the identifiability assumption, this implies $\bm{\theta}=\bm{\theta}_0$. Thus, we have proved that $D_{\rm GSM}(q_*,p_*)=0$ implies $\bm{\theta}=\bm{\theta}_0$. The converse is trivial.
\end{proof}

\begin{proof}[Proof of Corollary \ref{coro::wald test INID}]
By Conditions (C1)-(C4) and applying similar techniques to those used in the proof of Proposition \ref{prop::clt score function} in Section \ref{sect::tech details}, we obtain $\mathcal{I}_n^{-1}(\hat{\bm{\theta}})\mathcal{J}_n(\hat{\bm{\theta}})\mathcal{I}_n^{-1}(\hat{\bm{\theta}})\stackrel{p}{\longrightarrow}\mathcal{I}^{-1}(\bm{\theta}_0)\mathcal{J}(\bm{\theta}_0)\mathcal{I}^{-1}(\bm{\theta}_0).$ In addition, let $\bm{\Delta}=\left( I_\ell,\bm{0}_{\ell\times (p-\ell)} \right)^\top$ where $\bm{0}_{K_1\times K_2}$ denotes a $K_1\times K_2$ matrix with all elements being zeros. By Theorem \ref{thm::clt cts sm} and the continuous mapping theorem, we have, under the null hypothesis, $\sqrt{n}(\bm{\Delta}\hat{\bm{\theta}}-\bm{\theta}_{01})\stackrel{d}{\longrightarrow}N(\bm{0},\bm{\Delta} \mathcal{I}^{-1}(\bm{\theta}_0)\mathcal{J}(\bm{\theta}_0)\mathcal{I}^{-1}(\bm{\theta}_0)  \bm{\Delta}^\top)$. The above results, together with Slutsky's theorem and the continuous theorem, give
\[
T_w=(\bm{\Delta}\hat{\bm{\theta}}-\bm{\theta}_{01})^\top \left[ \bm{\Delta}\left\{ n^{-1} \mathcal{I}_n^{-1}(\hat{\bm{\theta}})\mathcal{J}_n(\hat{\bm{\theta}}) \mathcal{I}_n^{-1}(\hat{\bm{\theta}}) \right\}\bm{\Delta}^\top  \right]^{-1}(\bm{\Delta}\hat{\bm{\theta}}-\bm{\theta}_{01})\stackrel{d}{\longrightarrow}\chi^2_\ell.
\]
This completes the proof.

\end{proof}

\begin{proof}[Proof of Theorem \ref{thm::sm diff test}]
Under the null hypothesis, ${\rm H}_0:\bm{\theta}_1=\bm{\theta}_{01}$, we denote the resulting score matching estimation of $\bm{\theta}$ as $\widetilde{\bm{\theta}}$. Consider
\[
\mathcal{I}(\bm{\theta})=\begin{pmatrix} \mathcal{I}_{11}(\bm{\theta}) & \mathcal{I}_{12}(\bm{\theta}) \\
\mathcal{I}_{21}(\bm{\theta}) & \mathcal{I}_{22}(\bm{\theta})   \end{pmatrix},
\]
where $\mathcal{I}_{ij}(\bm{\theta})$ is the convergence of its corresponding information matrix with respect to $\bm{\theta}_i$ and $\bm{\theta}_j$ for $i,j\in\{1,2\}$. Then employing similar techniques to those used to deal with the Taylor series expansion of M-estimators (\cite{fan2001variable,zou2021network}), we obtain that
\[
\sqrt{n}(\widetilde{\bm{\theta}}-\bm{\theta}_0)=\sqrt{n}\widetilde{\mathcal{I}}_d(\bm{\theta}_0)\frac{\partial \hat{d}(\bm{\theta}_0)}{\partial \bm{\theta}}+o_p(1),
\]
where $\widetilde{\mathcal{I}}_d(\bm{\theta})=\begin{pmatrix} \bm{0}_{(p-l)\times {p-l}} & \bm{0}_{(p-l)\times p}\\\bm{0}_{p\times(p-l)} &\mathcal{I}_{22}^{-1}(\bm{\theta}) \end{pmatrix}$. This, together with the result of Theorem \ref{thm::clt cts sm}, implies that both $\bm{\theta}$ and $\widetilde{\bm{\theta}}$ are $\sqrt{n}$-consistent and 
\begin{align}\label{eq::proof thm sm diff test}
\sqrt{n}(\widetilde{\bm{\theta}}-\hat{\bm{\theta}})=\sqrt{n}\left\{ \widetilde{\mathcal{I}}_d(\bm{\theta}_0) - \mathcal{I}^{-1}(\bm{\theta}_0) \right\} \frac{\partial \hat{d}(\bm{\theta}_0)}{\partial \bm{\theta}} +o_p(1)=O_p(1).
\end{align}
Applying the Taylor series expansion, we have
\[
T_c=\sqrt{n}( \widetilde{\bm{\theta}}-\hat{\bm{\theta}} )^\top\left\{  -\frac{\partial^2 \hat{d} (\bar{\bm{\theta}}) }{\partial\bm{\theta}\partial\bm{\theta}^\top} \right\} \sqrt{n} ( \widetilde{\bm{\theta}}-\hat{\bm{\theta}} ),
\]
where $\bar{\bm{\theta}}$ lies between $\widetilde{\bm{\theta}}$ and $\hat{\bm{\theta}}$ and it is also $\sqrt{n}$-consistent. In addition, by Conditions (C1)-(C4) and applying similar techniques to those used in the proof of Proposition \ref{prop::clt score function} in Section \ref{sect::tech details}, we have
\[
- \frac{\partial^2 \hat{d} (\widetilde{\bm{\theta}}) }{\partial\bm{\theta}\partial\bm{\theta}^\top}\stackrel{p}{\longrightarrow} \mathcal{I}(\bm{\theta}_0).
\]
Therefore, $T_c=\sqrt{n} ( \widetilde{\bm{\theta}}-\hat{\bm{\theta}} )^\top\mathcal{I}(\bm{\theta}_0)\sqrt{n}( \widetilde{\bm{\theta}}-\hat{\bm{\theta}} )+o_p(1)$. By Proposition \ref{prop::clt score function} in Section \ref{sect::tech details}, we get
\[
\sqrt{n}\mathcal{K}^{-1/2}(\bm{\theta}_0)\frac{\partial \hat{d}(\bm{\theta}_0)}{\partial\bm{\theta}}\stackrel{d}{\longrightarrow} N(\bm{0},I_p).
\]
This, in conjunction with (\ref{eq::proof thm sm diff test}), leads to
\begin{align*}
T_c=&\left\{ \sqrt{n}\mathcal{K}^{-1/2}(\bm{\theta}_0) \frac{\partial \hat{d}(\bm{\theta}_0)}{\partial\bm{\theta}} \right\}^\top \mathcal{K}^{1/2}(\bm{\theta}_0) \{ \widetilde{\mathcal{I}}_d(\bm{\theta}_0) - \mathcal{I}^{-1}(\bm{\theta}_0) \} \mathcal{I}(\bm{\theta}_0)\\
&\times \{ \widetilde{\mathcal{I}}_d(\bm{\theta}_0) - \mathcal{I}^{-1}(\bm{\theta}_0) \} \mathcal{K}^{1/2}(\bm{\theta}_0) \left\{ \sqrt{n}\mathcal{K}^{-1/2}(\bm{\theta}_0) \frac{\partial \hat{d}(\bm{\theta}_0)}{\partial\bm{\theta}} \right\} +o_p(1).
\end{align*}
Using the fact that $\widetilde{\mathcal{I}}_d(\bm{\theta}_0)  \mathcal{I}(\bm{\theta}_0) \widetilde{\mathcal{I}}_d(\bm{\theta}_0) =\widetilde{\mathcal{I}}_d(\bm{\theta}_0)$ and $\{ \widetilde{\mathcal{I}}_d(\bm{\theta}_0) - \mathcal{I}^{-1}(\bm{\theta}_0) \} \mathcal{I}(\bm{\theta}_0)\{ \widetilde{\mathcal{I}}_d(\bm{\theta}_0) - \mathcal{I}^{-1}(\bm{\theta}_0) \} = \mathcal{I}^{-1}(\bm{\theta}_0)-\widetilde{\mathcal{I}}_d(\bm{\theta}_0)$, we further obtain
\begin{align*}
T_c=\left\{ \sqrt{n}\mathcal{K}^{-1/2}(\bm{\theta}_0) \frac{\partial \hat{d}(\bm{\theta}_0)}{\partial\bm{\theta}} \right\}^\top \mathcal{K}^{1/2}(\bm{\theta}_0) \{  \mathcal{I}^{-1}(\bm{\theta}_0)-\widetilde{\mathcal{I}}_d(\bm{\theta}_0)  \}\mathcal{K}^{1/2}(\bm{\theta}_0) \left\{ \sqrt{n}\mathcal{K}^{-1/2}(\bm{\theta}_0) \frac{\partial \hat{d}(\bm{\theta}_0)}{\partial\bm{\theta}} \right\} +o_p(1).
\end{align*}
Let $\lambda_1(\bm{\theta}_0),\ldots,\lambda_p(\bm{\theta}_0)$ be the eigenvalues of $\mathcal{K}^{1/2}(\bm{\theta}_0) \{  \mathcal{I}^{-1}(\bm{\theta}_0)-\widetilde{\mathcal{I}}_d(\bm{\theta}_0)  \}\mathcal{K}^{1/2}(\bm{\theta}_0)$. The above results, together with the continuous mapping theorem and Slutsky's, imply that $T_c$ follows a weighted chi-square distribution $\sum_{m=1}^p\lambda_m(\bm{\theta}_0)Z_m^2$ asymptotically.

Now we consider a change of parametrisation $\bm{\theta}_1^\prime=\bm{\theta}_1+\bm{U}\bm{\theta}_2$ and $\bm{\theta}_2^\prime=\bm{\theta}_2$ where $\bm{U}=-\mathcal{I}_{12}(\bm{\theta}_0)\mathcal{I}_{12}(\bm{\theta}_0)$. Note that $T_c$ is unchanged by this reparametrization since it is defined through minimization and this reparametrization constructs a pair of orthogonal parameters $\bm{\theta}_1^\prime$ and $\bm{\theta}_2^\prime$. After applying similar techniques to those used in the proof above, we could conclude that  $T_c$ follows a weighted chi-square distribution $\sum_{m=1}^\ell \lambda_m(\bm{\theta}_0)Z_m^2$ as $n \rightarrow \infty$, where $Z_1, \ldots , Z_m$ are independent $N(0,1)$ random variables and $\lambda_1(\bm{\theta}_0), \ldots , \lambda_{\ell}(\bm{\theta}_0)$ are the eigenvalues of the $\ell \times \ell$ matrix $\mathcal{A}^{1/2}(\bm{\theta}_0) \left \{\mathcal{I}_{11}(\bm{\theta}_0) - \mathcal{I}_{12}(\bm{\theta}_0)\mathcal{I}_{22}(\bm{\theta}_0)^{-1}\mathcal{I}_{21}(\bm{\theta}_0)  \right \}^{-1} \mathcal{A}^{1/2}(\bm{\theta}_0)$. This completes the entire proof.

\end{proof}

\begin{proof}[Proof of Theorem \ref{thm::wald test}]
We first show that the gradient of the score matching objective function in the vMF auto model is unbiased at the correct model. By the definition of score matching, $\bm{\theta}_0$ minimizes $d_{\rm SM}(\bm{\theta})$ and thus $\bm{W}\bm{\theta}_0=\bm{d}$. Note that the gradient of the score matching objective function in the vMF auto model gives ${\rm E}( \hat{\bm{W}}\bm{\theta}-\hat{\bm{d}} )=\bm{W}\bm{\theta}-\bm{d}$. It can be readily seen that the gradient of the score matching objective function in the vMF auto model is unbiased at the correct model.

Now, under the null hypothesis $H_0$, we have
\begin{align*}
f(\bm{y}_1,\ldots,\bm{y}_n|\bm{\theta}_0)&\propto\exp\left\{ \sum_{i=1}^n \bm{\beta}_0^\top \bm{y}_i\right\}=\prod_{i=1}^n \exp\left\{ \bm{\beta}_0^\top \bm{y}_i \right\},
\end{align*}
thus we can conclude that $\bm{y}_1,\ldots,\bm{y}_n$ are independent under the null hypothesis. By Proposition \ref{prop::quadratic clt} in Section \ref{sect::tech details}, we have $\hat{\bm{d}}\stackrel{p}{\longrightarrow} \bm{d}$. After applying similar techniques to those used in the proof of Proposition \ref{prop::quadratic clt} in Section \ref{sect::tech details}, we obtain $\hat{\bm{W}}\stackrel{p}{\longrightarrow} \bm{W}$. Recall that by the definition of score matching, $\bm{W}\bm{\theta}_0=\bm{d}$. Then by the CLT for martingale difference arrays in Proposition \ref{prop::quadratic clt} in Section \ref{sect::tech details}, we get $\sqrt{n}(\hat{\bm{d}}-\hat{\bm{W}}\bm{\theta}_0)\stackrel{d}{\longrightarrow} N(\bm{0},\mathcal{K}(\bm{\theta}_0))$, where $\mathcal{K}(\bm{\theta}_0)=\lim_{n\to\infty} n {\rm E}\left\{ \left( \hat{\bm{W}}\bm{\theta}_0-\hat{\bm{d}} \right)\left( \hat{\bm{W}}\bm{\theta}_0-\hat{\bm{d}} \right)^\top \right\}$. By Slutsky's theorem,
\[
\sqrt{n}(\hat{\bm{\theta}}-\bm{\theta}_0) \stackrel{d}{\longrightarrow} N (\bm{0},\bm{W}^{-1}\mathcal{K}(\bm{\theta}_0)\bm{W}^{-1}).
\]
This result, together with the continuous mapping theorem, implies that
\[
\sqrt{n}(\bm{\Delta}\hat{\bm{\theta}}) \stackrel{d}{\longrightarrow} N (0,\bm{\Delta}\bm{W}^{-1}\mathcal{K}(\bm{\theta}_0)\bm{W}^{-1}\bm{\Delta}).
\]
Employing similar techniques to those used in the proof of Proposition \ref{prop::quadratic clt} in Section \ref{sect::tech details}, we then have 
\[
\hat{\bm{W}}^{-1}\hat{\mathcal{K}}_n(\bm{\theta}_0)\hat{\bm{W}}^{-1} \stackrel{p}{\longrightarrow}  \bm{W}^{-1}\mathcal{K}(\bm{\theta}_0)\bm{W}^{-1}.
\]
The above results, together with Slutsky's theorem and the continuous mapping theorem, imply
\[
T_w=(\bm{\Delta}\hat{\bm{\theta}})^\top \left[ \bm{\Delta}\left\{  n^{-1}\hat{\bm{W}}^{-1}\mathcal{K}_n(\hat{\bm{\theta}})\hat{\bm{W}}^{-1} \right\}\bm{\Delta}^\top  \right]^{-1} \bm{\Delta}\hat{\bm{\theta}}\stackrel{d}{\longrightarrow}\chi^2(1).
\]

\end{proof}

\section*{Acknowledgments}

This work was partially supported by ANU PhD scholarship from the Australian National University. The work of Janice L. Scealy and Andrew T. A. Wood was supported by Australian Research Council grant DP220102232.

\section*{}

\newpage
\smallskip
\clearpage

\begin{center}
{\large\bf SUPPLEMENTARY MATERIAL}\\
\vskip 0.1truein
{\bf for}\\
\vskip 0.1truein
{\large \bf Generalized Score Matching}\\
\vskip 0.1truein
{\bf by}\\
\vskip 0.1truein
{ \bf Jiazhen Xu, \hskip 0.1truein Janice L. Scealy, \hskip 0.1truein Andrew T. A. Wood
\hskip 0.1truein and Tao Zou}
\end{center}

The supplementary material consists of a technical lemma, simulation and empirical results for the truncated Gaussian regression model, the CMP regression model and the novel auto model, generalized score matching for multivariate ordinal data and derivatives of proposed score matching objective functions.

This supplementary material consists of five parts. Section \ref{supp sect::tech lemma} introduces a technical lemma. Section \ref{supp sect::simulation} presents additional numerical results of the truncated Gaussian regression model, the CMP regression model and the vMF auto model. Section \ref{sup sec::gsm multi discrete} discusses generalized score matching for multivariate ordinal data. Section \ref{sup sec::score hessian} presents the detailed derivatives of the score matching objective functions for truncated Gaussian regression models and CMP regression models.

\setcounter{section}{0}

\makeatletter
\renewcommand\thesection{S\@arabic\c@section}
\renewcommand\thetable{S\@arabic\c@table}
\renewcommand \thefigure{S\@arabic\c@figure}
\makeatother

\section{Technical Lemma}\label{supp sect::tech lemma}

\begin{lemma}[\cite{DUNG2020108671}]\label{vector mda clt}
For a sequence of positive integers $\{k_n:n\geq 1\}$ such that $k_n\to\infty$ as $n\to\infty$, let $\{\bm{z}_{n,i}:1\leq i\leq k_n,n\geq 1\}$ be an array of $d$-dimensional martingale difference random vectors adapted to the filtration $\{\mathcal{F}_{n,i}:0\leq i\leq k_n,n\geq 1\}$ such that ${\rm E}(\|\bm{z}_{n,i}\|^2)<\infty$ for all $1\leq i\leq k_n$, $n\geq 1$. If
\begin{align}\label{vector mda clt condition1}
\sum_{i=1}^{k_n}{\rm E}\left(\| \bm{z}_{n,i} \|^2\mathbb{1}(\| \bm{z}_{n,i} \|>\epsilon) \Big| \mathcal{F}_{n,i-1}  \right) \stackrel{p}{\longrightarrow} 0
\end{align}
as $n\to\infty$ for each $\epsilon>0$, and
\begin{align}\label{vector mda clt condition2}
\sum_{i=1}^{k_n}{\rm E}\left(  \bm{z}_{n,i}\bm{z}_{n,i}^\top \Big| \mathcal{F}_{n,i-1} \right) \stackrel{p}{\longrightarrow} I_d
\end{align}
as $n\to\infty$, then
\[
\bm{S}_n=\sum_{i=1}^{k_n} \bm{z}_{n,i} \stackrel{d}{\longrightarrow} N(\bm{0},I_d)
\]
as $n\to\infty$.
\end{lemma}

It is readily seen that a sufficient condition for (\ref{vector mda clt condition1}) is that
\[
\sum_{i=1}^{k_n}{\rm E}\left(\| \bm{z}_{n,i} \|^{2+\delta} \Big| \mathcal{F}_{n,i-1}  \right) \stackrel{p}{\longrightarrow} 0
\]
for some $\delta>0$. In turn, applying Chebychev's inequality, it follows a sufficient condition for the latter condition is that 
\begin{align}\label{vector mda clt condition1 suff condition}
\sum_{i=1}^{k_n}{\rm E}\left\{ {\rm E}\left[\| \bm{z}_{n,i} \|^{2+\delta} \Big| \mathcal{F}_{n,i-1}  \right] \right\} \to 0
\end{align}

\section{Additional Simulation and Empirical Results}\label{supp sect::simulation}
This section presents additional numerical results of the truncated Gaussian regression model, the CMP regression model and the vMF auto model.
\subsection{Truncated Gaussian Regression Model}\label{supp subsect::simulation truncated gaussian}

We conducted a numerical study to evaluate the performance of the score matching estimator for a truncated Gaussian regression model. We are particularly interested in examining the bias, standard deviation and the root mean squared error of the score matching estimator.

In the setting of the simulation, the sample size $n$ varied in $\{200,500,1000\}$. Additionally, all simulations were conducted via $1000$ replicates. For the purpose of assessing the performance of parameter estimators, we denote $\hat{\bm{\theta}}^{(k)}$ as the vector estimation of $\bm{\theta}$ in the $k$-th replicate. For each component of $\bm{\theta}$, which is $\theta_j$, the averaged bias of $\hat{\theta}_j^{(k)}$, $k\in\{1,\ldots,1000\}$, is $\textrm{BIAS}=\frac{1}{1000}\sum_{k}(\hat{\theta}_j^{(k)}-\theta_j)$, and the standard deviation of $\hat{\theta}_j^{(k)}$ is $\textrm{SD}=\Big\{\frac{1}{1000}\sum_{k_1}(\hat{\theta}_j^{(k_1)}-\frac{1}{1000}\sum_{k_2}\hat{\theta}_j^{(k_2)})^2\Big\}^\frac{1}{2}$. Therefore, the root mean squared error is $\textrm{RMSE}=\sqrt{\textrm{SD}^2+\textrm{BIAS}^2}$. To compare SD with the asymptotic standard deviation of the estimators, we consider a sample version of the asymptotic standard deviation which is a consistent estimation of the intractable asymptotic standard deviation. This sample version is denoted by ASD and the details are carefully discussed in Section \ref{sect:: thm property}. Furthermore, we compared score matching with approximate MLE method. To calculate the approximate MLE of the truncated Gaussian regression model, we use the function \code{pmvnorm} from the \textsf{R} package \texttt{mvtnorm} (\cite{wilhelm2010tmvtnorm}) to approximate the normalizing constant. We use BIAS(SM) and BIAS(AMLE) to denote the average bias of the generalized score matching estimator and the approximate MLE, respectively. Similarly, RMSE(SM) and RMSE(AMLE) are used to denote the root mean squared error of the generalized score matching estimator and the approximate MLE, respectively.

\begin{figure}
\begin{center}
\includegraphics[scale=0.4]{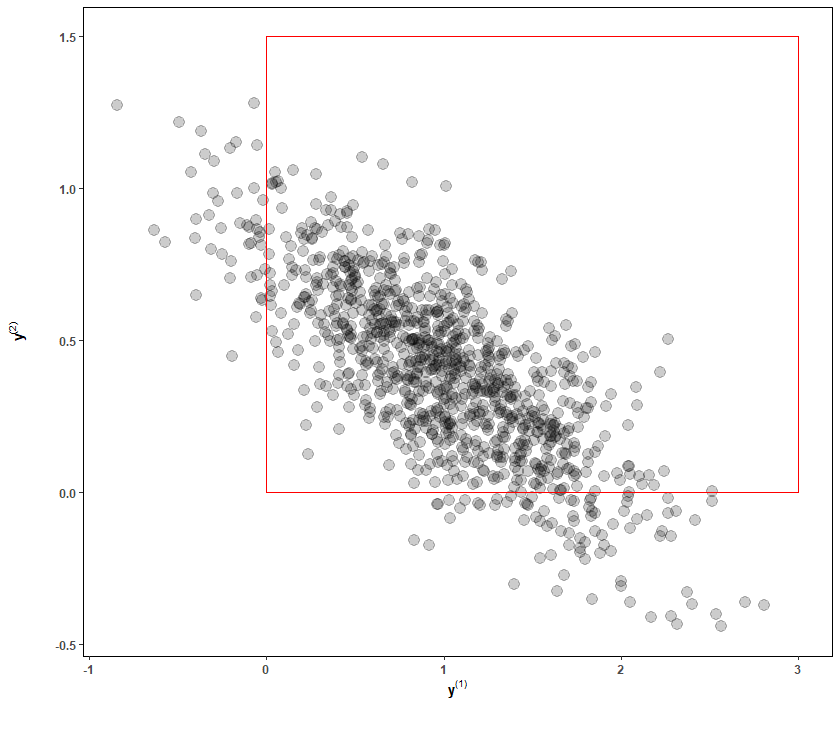}
\end{center}
\caption{Random samples from a truncated Gaussian regression model generated by the rejection algorithm with $\bm{y}^{(1)}=(y_{11},\ldots,y_{n1})^\top$ and $\bm{y}^{(2)}=(y_{12},\ldots,y_{n2})^\top$\label{fig:first}}
\end{figure}

We simulated data from a truncated Gaussian regression model as follows. For $i\in\{1,\ldots,n\}$, consider the $2\times 1$ covariate vector $\bm{x}_i$ with $\bm{x}_i=(x_{i1},x_{i2})^\top$, $x_{i1}\equiv 1$, $x_{i2}$ being independent and identically generated from the standard normal distribution $N(0,1)$, and their corresponding regression parameters are ${\rm vec}(\bm{B}_0)=(B_{11},B_{21},B_{12},B_{22})^\top=(1,0.4,-0.5,0.2)^\top$. The true precision matrix $\bm{\Lambda}_0$ is set to be ${\rm vec}(\bm{\Lambda}_0)=(\Lambda_{11},\Lambda_{21},\Lambda_{12},\Lambda_{22})^\top$ $=(20,10,10,30)^\top$. It is worth noting that the covariate matrix is fixed across the replications. The domain of the response $\bm{y}_i=(y_{i1},y_{i2})^\top$ is chosen to be $\mathbb{R}_{>0}^2$ where $\mathbb{R}_{>0}$ denotes the set of positive real numbers. We used the rejection algorithm (\cite{wilhelm2010tmvtnorm}) to generate the data. That is, we continued generating candidate $\bm{y}_i$ from the bivariate Gaussian distribution $N(\bm{B}_0\bm{x}_i,\bm{\Lambda}_0^{-1})$ until the candidate was located inside the support region $\mathbb{R}_{>0}^2$. Figure \ref{fig:first} shows random samples from the bivariate Gaussian distribution $N(\bm{B}_0\bm{x}_i,\bm{\Lambda}_0^{-1})$ and the samples from the truncated Gaussian regression model are located inside the red box.

Our parameter of interest is $\bm{\theta}=\left({\rm vec}^\top(\bm{B}),{\rm vech}^\top(\bm{\Lambda})\right)^\top$. If we conduct a log transformation to the response, then we will have a new variable $\widetilde{\bm{y}}_i=({\rm log} y_{i1},{\rm log} y_{i2})^\top$. After the transformation, the domain of the response $\widetilde{\bm{y}}_i$ is $\mathbb{R}^2$ and the log-transformed truncated Gaussian regression model is given by
\[
p(\widetilde{\bm{y}}_i|\bm{x}_i,\bm{\theta})=\frac{1}{Z_i(\bm{\theta})}\exp \left\{ \widetilde{\bm{y}}_i^\top \bm{l}-\frac{1}{2}\left(\bm{t}(\widetilde{\bm{y}}_i)-\bm{B}\bm{x}_i \right)^\top\bm{\Lambda} \left(\bm{t}(\widetilde{\bm{y}}_i)-\bm{B}\bm{x}_i  \right) \right\},
\]
where $\bm{l}=(1,1)^\top$ and $\bm{t}(\widetilde{\bm{y}}_i)=(e^{\widetilde{y}_{i1}},e^{\widetilde{y}_{i2}})^\top$. The score matching objective function is then in the form of
\begin{align}\label{log transformed tgm objective function}
\hat{d}_{\rm SM}(\bm{\theta})=&\frac{1}{n}\sum_{i=1}^n\Bigg\{ -4 \left(\bm{t}(\widetilde{\bm{y}}_i)-\bm{B}\bm{x}_i  \right)^\top\bm{\Lambda} \bm{t}(\widetilde{\bm{y}}_i)-2{\rm tr}\left(\bm{T}(\widetilde{\bm{y}}_i)\bm{\Lambda} \bm{T}(\widetilde{\bm{y}}_i)\right) \nonumber \\
&+\left(\bm{t}(\widetilde{\bm{y}}_i)-\bm{B}\bm{x}_i  \right)^\top \bm{\Lambda} \bm{T}(\widetilde{\bm{y}}_i) \bm{T}(\widetilde{\bm{y}}_i) \bm{\Lambda} \left(\bm{t}(\widetilde{\bm{y}}_i)-\bm{B}\bm{x}_i \right) \Bigg\},
\end{align}
where $\bm{T}(\widetilde{\bm{y}}_i)={\rm diag}(e^{\widetilde{y}_{i1}},e^{\widetilde{y}_{i2}})$. 

We next show that the objective function (\ref{log transformed tgm objective function}) of the log-transformed truncated Gaussian regression model is equivalent to the objective function of the truncated Gaussian regression model using weight functions (\cite{yu2019generalized}). Based on \cite{yu2019generalized}, the empirical objective function of the truncated Gaussian regression model can be represented by
\begin{align}\label{tgm objective function}
\hat{d}_{\rm TSM}(\bm{\theta})=&\frac{1}{n}\sum_{i=1}^n\Bigg\{ 2\bm{g}^\prime(\bm{y}_i)^\top\frac{\partial \log p(\bm{y}_i)}{\partial \bm{y}_i}+\left(  \bm{g}^{\frac{1}{2}}(\bm{y}_i)\odot\frac{\partial \log p(\bm{y}_i)}{\partial \bm{y}_i} \right)^\top \left(  \bm{g}^{\frac{1}{2}}(\bm{y}_i)\odot\frac{\partial \log p(\bm{y}_i)}{\partial \bm{y}_i} \right)\nonumber \\
&+  2{\rm tr}\left[ \left( \bm{g}^{\frac{1}{2}}(\bm{y}_i)\bm{g}^{\frac{1}{2}}(\bm{y}_i)^\top \right)\odot  \frac{\partial^2 \log p(\bm{y}_i)}{\partial \bm{y}_i \partial \bm{y}_i^\top}  \right]  \Bigg\}\nonumber\\
=&\frac{1}{n}\sum_{i=1}^n\Bigg\{ -2\bm{g}^\prime(\bm{y}_i)^\top \bm{\Lambda} (\bm{y}_i-\bm{B}\bm{x}_i )\nonumber\\
& + \left[ \bm{g}^{\frac{1}{2}}(\bm{y}_i)\odot \bm{\Lambda}  (\bm{y}_i-\bm{B}\bm{x}_i ) \right]^\top  \left[ \bm{g}^{\frac{1}{2}}(\bm{y}_i)\odot \bm{\Lambda}  (\bm{y}_i-\bm{B}\bm{x}_i ) \right]  \nonumber \\
&- 2{\rm tr}\left[ \left( \bm{g}^{\frac{1}{2}}(\bm{y}_i)\bm{g}^{\frac{1}{2}}(\bm{y}_i)^\top \right)\odot  \Lambda  \right]  \Bigg\} \nonumber\\
=&\frac{1}{n}\sum_{i=1}^n\Bigg\{ -2\bm{g}^\prime(\bm{y}_i)^\top \bm{\Lambda} (\bm{y}_i-\bm{B}\bm{x}_i ) + (\bm{y}_i-\bm{B}\bm{x}_i )^\top \bm{\Lambda}   \bm{G}^{\frac{1}{2}}(\bm{y}_i) \bm{G}^{\frac{1}{2}}(\bm{y}_i) \Lambda (\bm{y}_i-\bm{B}\bm{x}_i ) \nonumber \\
&- 2{\rm tr}\left( \bm{G}^{\frac{1}{2}}(\bm{y}_i)  \Lambda \bm{G}^{\frac{1}{2}}(\bm{y}_i)\right)  \Bigg\},
\end{align}
where $\bm{g}(\bm{y}_i)=(g_1(y_{i1}),g_1(y_{i2}))^\top$ is a vector of weight functions, $\bm{g}^\prime(\bm{y}_i)=(\frac{\partial g_1(y_{i1})}{\partial y_{i1}},\frac{\partial g_2(y_{i1})}{\partial y_{ij}})^\top$, $\bm{g}^{\frac{1}{2}}(\bm{y}_i)=(\sqrt{g_1(y_{i1})},\sqrt{g_1(y_{i2})})^\top$, $\bm{G}^{\frac{1}{2}}(\bm{y}_i)={\rm diag}\left(  \bm{g}^{\frac{1}{2}}(\bm{y}_i) \right)$ and $\odot$ denotes the Hadamard product. It can be readily seen that the objective function (\ref{log transformed tgm objective function}) of the log-transformed truncated Gaussian regression model is equivalent to the objective function (\ref{tgm objective function}) of the truncated Gaussian regression model when we set the weight function to be $\bm{g}(\bm{y}_i)=(y_{i1}^2,y_{i2}^2)^\top$.

For the truncated Gaussian regression model, Table \ref{tnormal:sim2} below reports the BIAS(SM), SD, ASD and RMSE(SM) of the score matching estimator, along with the BIAS(AMLE) and RMSE(AMLE) of the approximate MLE, via $1000$ replications with three sample sizes. According to Table \ref{tnormal:sim2}, we find that the absolute values of BIAS(SM) and SD generally become smaller for all parameter estimates as $n$ becomes larger. It is not surprising that RMSE(SM) shows the same pattern. Furthermore, we notice that the absolute values of the difference between SD and ASD also become smaller for all estimators when $n$ gets larger. The above findings support our theoretical results that the score matching estimator for continuous data is consistent and asymptotically normal.

Additionally, three points should be noted. First, the bias for the score matching approach decreases as the sample size increases while for the approximate MLE the bias seems to stay nearly constant and non-zero, suggesting that the MLE is not being computed with sufficient accuracy to be consistent. Second, the SDs of both the score matching estimators and the approximate MLEs decrease, as expected, when the sample size increases. Note that the SDs of the approximate MLEs are typically smaller than the SDs of the score matching estimators; typically the SD ratio is above 80\%. Assuming that the approximation in the calculation of the MLEs mainly affects bias and not the SD, it is reasonable to suppose that the SD ratio gives a reasonable approximation of the efficiency of the score matching estimators relative to the (exact but unobserved) MLEs. Thus it is reasonable to suppose that the efficiency is typically above 80\% which indicates that the loss of efficiency in using the score matching estimators relative to the exact MLEs is fairly modest. Third, the RMSEs tend to
be a lot smaller for the score matching estimators of the $B_{jk}$ compared with the RMSE of the approximate MLEs. In contrast the RMSEs for the $\Lambda_{jk}$ tends to be smaller for the approximate MLE than the score matching RMSEs for smaller sample sizes but the reverse is usually true for larger sample sizes. Overall, in terms of the RMSE, the score matching estimators are competitive with (and typically superior to, especially with larger sample sizes) the approximate MLEs.

\begin{table}[H]
\caption{Comparison of the score matching estimation and approximate MLEs of the parameters ($B_{11}=1$, $B_{21}=0.4$, $B_{12}=-0.5$, $B_{22}=0.2$, $\Lambda_{11}=20$, $\Lambda_{12}=10$, $\Lambda_{22}=30$) for the truncated Gaussian regression model. Six measures are considered: the averaged bias of the estimate (BIAS(SM)), the compared average bias of the estimate calculated by approximate MLE, the standard deviation of the estimate (SD), the theoretical standard deviation of the estimate in the estimation theory (ASD), the root mean squared error of the estimate (RMSE(SM)), and the compared root mean squared error of the estimate calculated by approximate MLE.\label{tnormal:sim2}}
\centering
\scalebox{0.8}{
\begin{tabular}{lclrrrrrrr}
  \hline
&$n$&Measure& $\hat{B}_{11}$ & $\hat{B}_{21}$ & $\hat{B}_{12}$ & $\hat{B}_{22}$ & $\hat{\Lambda}_{11}$ & $\hat{\Lambda}_{12}$ &  $\hat{\Lambda}_{22}$ \\ 
  \hline
&$n=200$&BIAS(SM) & 0.0015 & -0.0010 & 0.0006 & -0.0007 & 1.0372 & 0.5327 & 1.7498 \\ 
&&BIAS(AMLE)  & 0.2146 & -0.0705 & 0.0846 & -0.0289 & 1.9209 & -0.7747 & -0.0800 \\ 
    &&SD   & 0.0313 & 0.0325 & 0.0367 & 0.0329 & 2.9603 & 2.4360 & 5.5817 \\   
     &&ASD& 0.0281 & 0.0255 & 0.0322 & 0.0270 & 2.6505 & 3.2547 & 4.4425 \\ 
    &&RMSE(SM)  & 0.0314 & 0.0325 & 0.0367 & 0.0329 & 3.1368 & 2.4936 & 5.8495 \\
    &&RMSE(AMLE) & 0.2153 & 0.0727 & 0.0867 & 0.0342 & 2.5391 & 1.3443 & 1.9827 \\ 
   \hline
  &$n=500$&BIAS(SM)& 0.0010 & -0.0006 & 0.0005 & -0.0001 & 0.4147 & 0.0982 & 0.8257 \\ 
   &&BIAS(AMLE)  & 0.2146 & -0.0704 & 0.0805 & -0.0281 & 1.3592 & -0.6840 & 0.1725 \\ 
     &&SD  & 0.0188 & 0.0189 & 0.0240 & 0.0221 & 1.9426 & 1.5699 & 3.4566 \\ 
     &&ASD   & 0.0177 & 0.0160 & 0.0216 & 0.0183 & 1.7331 & 2.0596 & 2.8432 \\  
    &&RMSE(SM) & 0.0177 & 0.0160 & 0.0216 & 0.0183 & 1.7331 & 2.6596 & 2.8432 \\  
    &&RMSE(AMLE) & 0.2149 & 0.0713 & 0.0813 & 0.0303 & 1.7058 & 1.1913 & 1.4017 \\ 
   \hline
   &$n=1000$&BIAS(SM) & -0.0001 & -0.0002 & -0.0000 & -0.0002 & 0.2134 & 0.1025 & 0.3003 \\  
    &&BIAS(AMLE) & 0.2144 & -0.0711 & 0.0829 & -0.0290 & 1.2548 & -1.0830 & -0.1154 \\ 
    &&SD  & 0.0139 & 0.0148 & 0.0180 & 0.0154 & 1.2424 & 1.1253 & 2.4593 \\  
    &&ASD& 0.0131 & 0.0125 & 0.0156 & 0.0128 & 1.2441 & 1.5766 & 2.0818 \\  
    &&RMSE(SM)  & 0.0139 & 0.0148 & 0.0180 & 0.0154 & 1.2606 & 1.1299 & 2.4776 \\
    &&RMSE(AMLE)  & 0.2146 & 0.0715 & 0.0833 & 0.0301 & 1.5371 & 1.3049 & 1.0130 \\ 
   \hline
\end{tabular}}
\end{table}

\begin{figure}
\begin{center}
\includegraphics[scale=0.7]{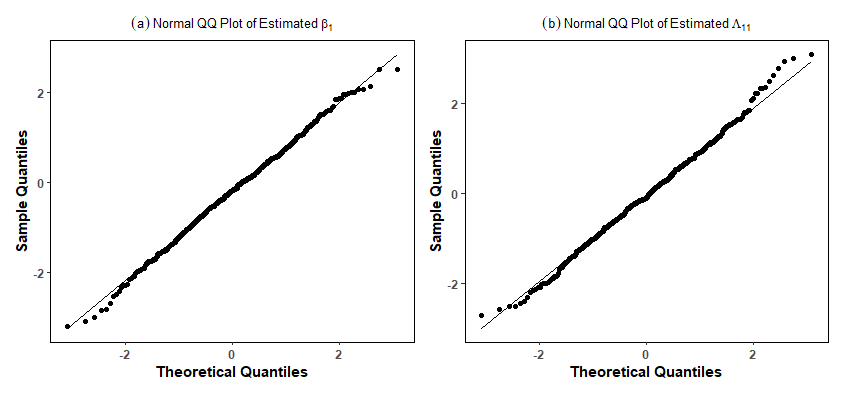}
\end{center}
\caption{Normal QQ plot of (a) $\hat{\beta}_1$ and (b) $\hat{\Lambda}_{11}$ for the truncated Gaussian regression model.\label{fig:qqplot}}
\end{figure}

To illustrate the asymptotic normality of the score matching estimation, Figure \ref{fig:qqplot} shows the QQ plot of $\hat{\beta}_1$ and $\hat{\Lambda}_{11}$ for $n=1000$, which also supports our theoretical results. 

We next assess the finite sample performance of the score-matching-based Wald test and the change-in-score-matching test via evaluating the empirical size with the significance levels ranging from 0.01 to 0.30 and the empirical power with the significance level 0.05. The empirical size and power are the percentages of rejections under ${\rm H}_0$ and ${\rm H}_1$, respectively via the hypothesis test
\[
{\rm H}_0: (B_{12},B_{22})^\top=\bm{0}~~{\rm versus}~~{\rm H}_1:\widetilde{\bm{\beta}}\neq\bm{0},
\]
with $1000$ realizations. Note that this hypothesis testing can be represented as the one discussed in Section \ref{sect:: thm property} by rearranging $\bm{\theta}$. The empirical size is the percentage of rejections under the setting of $(B_{11},B_{21},B_{12},B_{22})=(1,0.4,0,0)$ and ${\rm vec}(\bm{\Lambda}_0)=(\Lambda_{11},\Lambda_{21},\Lambda_{12},\Lambda_{22})^\top$ $=(20,10,10,30)^\top$. The empirical power is the percentage of rejections under the setting of $(B_{11},B_{21},B_{12},B_{22})=(1,0.4,-0.5\iota,0.2\iota)$ and ${\rm vec}(\bm{\Lambda}_0)=(\Lambda_{11},\Lambda_{21},\Lambda_{12},\Lambda_{22})^\top$ $=(20,10,10,30)^\top$, where the signal strength $\iota>0$. 

\begin{figure}
\centering
\includegraphics[width=1\linewidth]{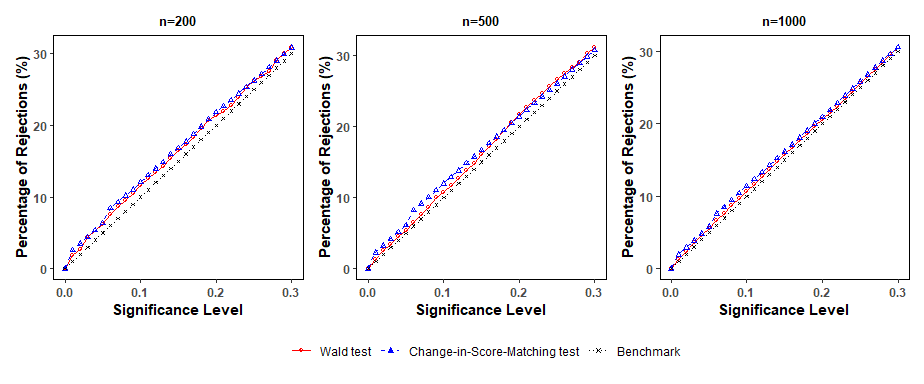}
\caption{The empirical sizes of the score-matching-based Wald test and the change-in-score-matching test for the significance levels ranging from 0.01 to 0.30 under the setting of the truncated Gaussian regression model. The benchmark represents the ideal case when the percentage of rejections from 1000 replications is equal to the significance level.}
\label{fig::simulation tgr test compare size}
\end{figure}

\begin{figure}
\centering
\includegraphics[width=1\linewidth]{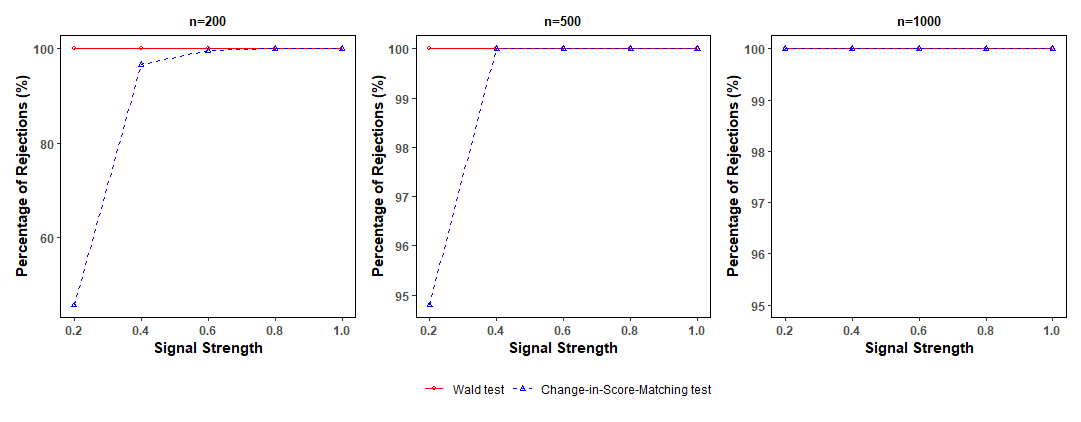}
\caption{The empirical power of the score-matching-based Wald test and the change-in-score-matching test at a nominal level of 0.05. The signal strengths $\iota=0.2,0.4,0.6,0.8$ and 1 which correspond to the settings $(B_{11},B_{21},B_{12},B_{22})=(1,0.4,-0.5\iota,0.2\iota)$, respectively.}
\label{fig::simulation tgr test compare power}
\end{figure}

Figure \ref{fig::simulation tgr test compare size} shows that the empirical sizes of the score-matching-based Wald test and the change-in-score-matching test are almost identical to the predetermined significance levels as $n=1000$. Figure \ref{fig::simulation tgr test compare power} shows the empirical powers of these two tests tend to 100\% when the sample size $n$ or the signal strength $\iota$ gets larger. However, we find that the change-in-score-matching test is not powerful when the signal strength $\iota$ is small. These findings indicate that these two tests perform well when $n$ is large and the score-matching-based Wald test performs much better when $n$ is small.

\subsection{CMP Regression Model}\label{supp subsect::simulation cmp}
\begin{figure}
    \centering
    \includegraphics[scale=0.3]{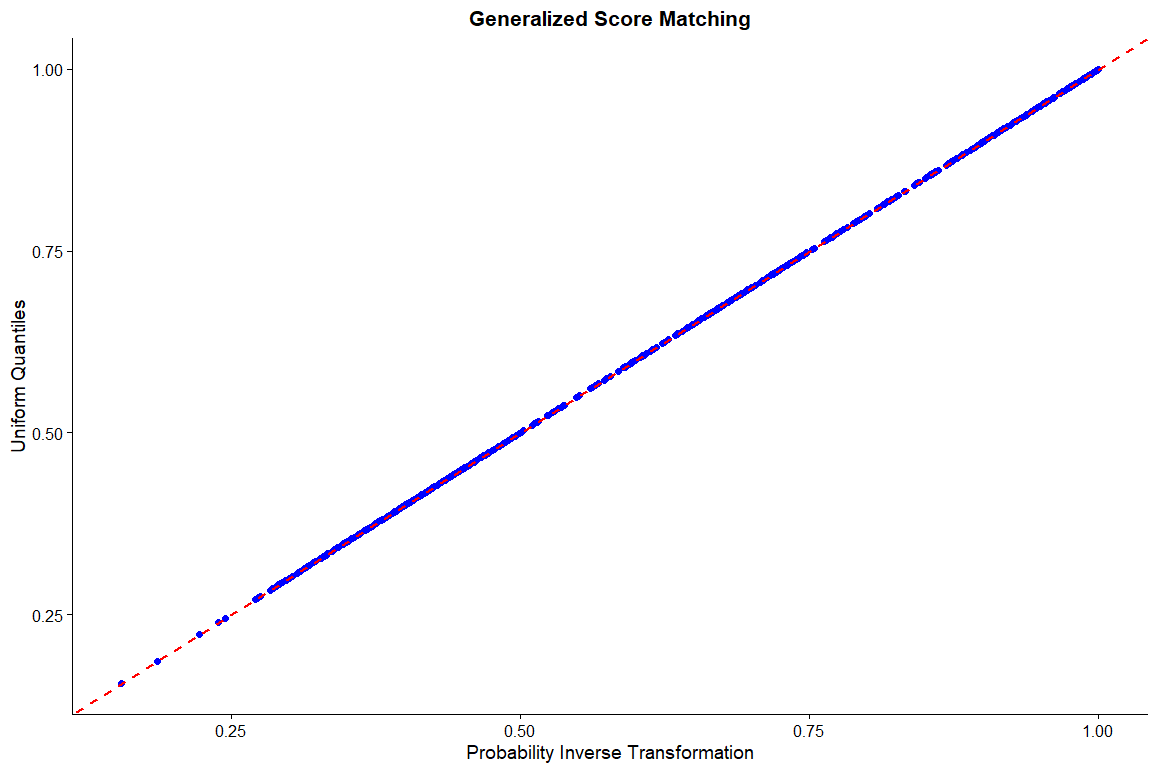} % second figure itself
    \caption{PIT-uniform quantile plot for the fitted CMP regression model by generalized score matching.\label{fig::case study}}
\end{figure}

Figure \ref{fig::case study} is a PIT-uniform quantile plot. It shows reasonable closeness to uniformity, which indicates that the fitted CMP regression model using generalized score matching is appropriate.

\begin{table}[htbp!]
\caption{Comparison of the generalized score matching estimations and approximate MLEs of the parameters ($\beta_1=-0.3141$, $\beta_2=-0.0893$, $\beta_3=0.0445$, $\beta_4=-0.0705$, $\beta_5=0.0693$, $\beta_6=0.0830$ $\nu=0.2564$) for the CMP regression model. Six measures are considered: the averaged bias of the estimate (BIAS(SM)), the compared average bias of the estimate calculated by \code{glm.cmp} (BIAS(AMLE)), the standard deviation of the estimate (SD), the theoretical standard deviation of the estimate in the estimation theory (ASD), the root mean squared error of the estimate (RMSE(SM)), and the compared root mean squared error of the estimate calculated by \code{glm.cmp} (RMSE(AMLE)).\label{normal:sim4}}
\centering
\scalebox{0.8}{
\begin{tabular}{lclrrrrrrr}
  \hline
&$n$&Measure& $\hat{\beta}_1$ & $\hat{\beta}_2$ & $\hat{\beta}_3$ & $\hat{\beta}_4$ & $\hat{\beta}_5$ & $\hat{\beta}_6$ & $\hat{\nu}$ \\ 
  \hline
&$n=200$&BIAS(SM)& 0.0204 & 0.0053  & 0.0084 & -0.0051 & 0.0063& 0.0086 & 0.0643 \\
  &&BIAS(AMLE)  & 0.1986 & 0.0713 & 0.0211 & 0.0154 & 0.0292  & 0.0424 & 0.2950 \\
  &&SD  & 0.1970 & 0.1531 & 0.1598 & 0.0914 & 0.0743 & 0.0752 & 0.1491 \\
  &&ASD & 0.1871 & 0.1391 & 0.1466 & 0.0838 & 0.0700 & 0.0677 & 0.1528 \\
  &&RMSE(SM)& 0.1981 & 0.1532 & 0.1600 & 0.0915 & 0.0746 & 0.0756 & 0.1623 \\
  &&RMSE(AMLE) & 0.3005 & 0.1945 & 0.1831 & 0.0917 & 0.0817 & 0.0845 & 0.3175 \\
   \hline
  &$n=500$&BIAS(SM)& 0.0152 & -0.0050 & -0.0038 &  -0.0034 & 0.0060 & 0.0037 & 0.0391 \\
&&BIAS(AMLE)  & 0.2076 & 0.0517 & -0.0198 & 0.0123 & 0.0328 & 0.0227 & 0.2825 \\
 &&SD & 0.1248  & 0.0996 & 0.1055 & 0.0515 & 0.0466 & 0.0385 & 0.1047 \\
 &&ASD & 0.1170 & 0.0850 & 0.0944 & 0.0481 & 0.0419 & 0.0337 & 0.0963 \\ 
&&RMSE(SM) & 0.1257 & 0.0997 & 0.1055 & 0.0516& 0.0470 & 0.0387 & 0.1118 \\
&&RMSE(AMLE) & 0.2585 & 0.1432 & 0.1420 & 0.0659 & 0.0656 & 0.0519 & 0.2992 \\
   \hline
 &$n=1000$&BIAS(SM) & 0.0132 & 0.0032 & -0.0017 & -0.0019 & 0.0011 & 0.0013 & 0.0254 \\
&&BIAS(AMLE)  & 0.2153 & 0.0533 & -0.0061 & 0.0130 & 0.0347 & 0.0336& 0.2899 \\
 &&SD  & 0.0889 & 0.0695 & 0.0770 & 0.0392 & 0.0346 & 0.0274 & 0.0798 \\
&&ASD & 0.0819 & 0.0606 & 0.0669 & 0.0357 & 0.0302 & 0.0240 & 0.0746 \\  
&&RMSE(SM) & 0.0899 & 0.0695 & 0.0770 & 0.0393 & 0.0346 & 0.0274 & 0.0837 \\  
&&RMSE(AMLE) & 0.2563 & 0.1353 & 0.1266 & 0.0602 & 0.0617 & 0.0507 & 0.3041 \\
   \hline
\end{tabular}}
\end{table}

Table \ref{normal:sim4} reports the BIAS(SM), SD, ASD and RMSE(SM) of the generalized score matching estimator, along with the BIAS(AMLE) and RMSE(AMLE) of the approximate MLE, via $1000$ replications with three sample sizes. According to Table \ref{normal:sim4}, we find that the absolute values of BIAS(SM) and SD generally become smaller for all parameter estimates as $n$ becomes larger. It is not surprising that RMSE(SM) shows the same pattern. Furthermore, we notice that the absolute values of the difference between SD and ASD also become smaller for all estimators when $n$ gets larger. The above findings support our theoretical results that the generalized score matching estimator for ordinal data is consistent and asymptotically normal. Moreover, we notice that the generalized score matching method is much more accurate when the approximate MLE is biased.

\subsection{Auto Model for Spherical Data}

We conducted a numerical study to evaluate the performance of the score matching estimator for the vMF auto model. We are particularly interested in examining the bias, standard deviation and the root mean squared error of the score matching estimator.

In the setting of the simulation, the sample size $n$ varied in $\{200,500,1000\}$. Additionally, all simulations were conducted via $1000$ replicates. For the purpose of assessing the performance of parameter estimators, we denote $\hat{\bm{\theta}}^{(k)}$ as the vector estimation of $\bm{\theta}$ in the $k$-th replicate. For each component of $\bm{\theta}$, which is $\theta_j$, the averaged bias of $\hat{\theta}_j^{(k)}$, $k\in\{1,\ldots,1000\}$, is $\textrm{BIAS}=\frac{1}{1000}\sum_{k}(\hat{\theta}_j^{(k)}-\theta_j)$, and the standard deviation of $\hat{\theta}_j^{(k)}$ is $\textrm{SD}=\Big\{\frac{1}{1000}\sum_{k_1}(\hat{\theta}_j^{(k_1)}-\frac{1}{1000}\sum_{k_2}\hat{\theta}_j^{(k_2)})^2\Big\}^\frac{1}{2}$. Therefore, the root mean squared error is $\textrm{RMSE}=\sqrt{\textrm{SD}^2+\textrm{BIAS}^2}$.

We simulated data from the vMF auto model under the null hypothesis as follow. Note that under the null hypothesis, $\xi_0=0$ and thus for $i\in\{1,\ldots,n\}$, we independently generated $\bm{y}_i\in\mathcal{S}^5$ from the vMF distribution with the mean direction $\bm{\beta}_0/\|\bm{\beta}_0\|$ and the concentration $\|\bm{\beta}_0\|$, where $\bm{\beta}_0=(\beta_1,\beta_2,\beta_3,\beta_4,\beta_5,\beta_6)^\top=(2.8792,2.3916,1.9828,1.5974,6.5620,1.6320)^\top$. The neighborhood structure we considered is similar to the one obtained from the data set in Section \ref{sect::sm dependent}. Table \ref{sim results::auto model} below reports the BIAS(SM), SD, ASD and RMSE(SM) of the score matching estimator via $1000$ replications with three sample sizes. Table \ref{sim results::auto model} yields similar results to those obtained in the numerical studies for the CMP regression model and the truncated Gaussian regression model.

\begin{table}[H]
\caption{The BIAS, SD, and RMSE of the score matching estimator for $\bm{\beta}_0=(\beta_1,\beta_2,\beta_3,\beta_4,\beta_5,\beta_6)^\top=(2.8792,2.3916,1.9828,1.5974,6.5620,1.6320)^\top$ and $\xi_0=0$.\label{sim results::auto model}}

\begin{center}

\begin{tabular}{lclrrrrrrr}
  \hline
&$n$&Measure& $\hat{\beta}_1$ & $\hat{\beta}_2$ & $\hat{\beta}_3$ & $\hat{\beta}_4$ & $\hat{\beta}_5$ & $\hat{\beta}_6$ & $\hat{\xi}$  \\ 
  \hline
&$n=200$&BIAS& 0.2131 & 0.1721 & 0.1472 & 0.0933 & 0.4582 & 0.1089 & -0.0443 \\ 
  &&SD    & 0.4873 & 0.4320 & 0.3714 & 0.3389 & 0.9823 & 0.3515 & 0.0964 \\ 
  &&RMSE   & 0.5318 & 0.4650 & 0.3995 & 0.3515 & 1.0839 & 0.3680 & 0.1061 \\ 
   \hline
  &$n=500$&BIAS& 0.0786 & 0.0626 & 0.0566 & 0.0436 & 0.1852 & 0.0477 & -0.0170 \\ 
  &&SD   & 0.2915 & 0.2682 & 0.2274 & 0.2154 & 0.5867 & 0.2149 & 0.0581 \\ 
  &&RMSE & 0.3019 & 0.2754 & 0.2343 & 0.2198 & 0.6153 & 0.2201 & 0.0606 \\ 

   \hline
 &$n=1000$&BIAS& 0.0330 & 0.0304 & 0.0217 & 0.0217 & 0.0822 & 0.0170 & -0.0075 \\ 
  &&SD   & 0.2036 & 0.1819 & 0.1586 & 0.1459 & 0.4074 & 0.1507 & 0.0404 \\ 
  &&RMSE & 0.2062 & 0.1845 & 0.1601 & 0.1475 & 0.4157 & 0.1516 & 0.0411 \\ 
  
   \hline
\end{tabular}
\end{center}

\end{table}

\begin{figure}
\centering
\includegraphics[width=1\linewidth]{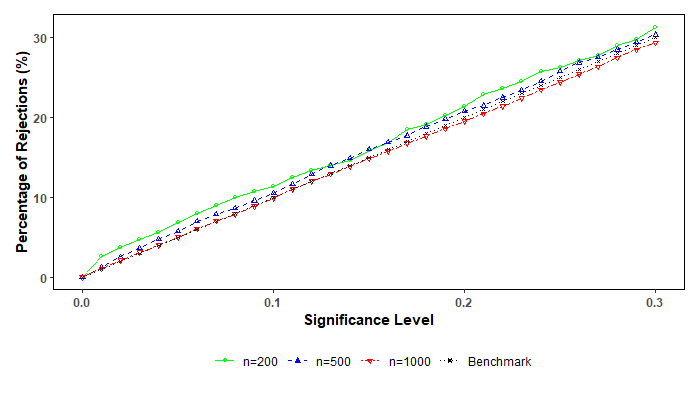}
\caption{The empirical sizes of the score-matching-based Wald test for the significance levels ranging from 0.01 to 0.30 under the setting of the vMF auto model for $n$ varies in $\{200,500,1000\}$. The benchmark represents the ideal case when the percentage of rejections from 1000 replications is equal to the significance level.}
\label{fig::simulation auto model test compare size}
\end{figure}

We next assess the finite sample performance of the score-matching-based Wald test via evaluating the empirical size with the significance levels ranging from 0.01 to 0.30. Figure \ref{fig::simulation auto model test compare size} shows that the empirical sizes of the score-matching-based Wald test is almost identical to the predetermined significance levels at $n=1000$.

\section{Generalized Score Matching for Multivariate Ordinal Data}\label{sup sec::gsm multi discrete}
In this section, we will present the theoretical properties of our proposed generalized score matching for INID multivariate ordinal data $\bm{y}$.

Recall that the overall population generalized score matching objective function is given by
\begin{align}\label{discrete population function multi}
D_{\rm GSM}(q_*,p_*)=&\frac{1}{n}\sum_{i=1}^n{\rm E} \Bigg(\sum_{j=1}^d \Bigg\{\left[ t\left( \frac{p_i(\bm{y}_i^{(j_{+})}|\bm{\theta})}{p_i(\bm{y}_i|\bm{\theta})} \right) - t\left( \frac{q_i(\bm{y}_i^{(j_{+})})}{q_i(\bm{y}_i)} \right) \right]^2 \nonumber\\
&+ \left[ t\left( \frac{p_i(\bm{y}_i|\bm{\theta})}{p_i(\bm{y}_i^{(j_{-})}|\bm{\theta})}\right) - t\left( \frac{q_i(\bm{y}_i)}{q_i(\bm{y}_i^{(j_{-})})} \right) \right]^2 \Bigg\}\Bigg).
\end{align}
The following theorem will show that (\ref{discrete population function multi}) is tractable. The proofs of all the following theorems and propositions are similar to those in the univariate case and are omitted.
\begin{theorem}\label{thm:: tractable obj funct for GSM multi}
(\ref{discrete population function multi}) can be decomposed as
\[
D_{\rm GSM}(q_*,p_*)=g(q_*)+d_{\rm GSM}(q_*,p_*),
\]
where
\[
d_{\rm GSM}(q_*,p_*)=\frac{1}{n}\sum_{i=1}^n {\rm E}\Bigg(\sum_{j=1}^d\Bigg\{ t\left( \frac{p_i(\bm{y}_i^{(j_{+})}|\bm{\theta})}{p_i(\bm{y}_i|\bm{\theta})} \right)^2 + t\left( \frac{p_i(\bm{y}_i|\bm{\theta})}{p_i(\bm{y}_i^{(j_{-})}|\bm{\theta})}\right)^2-2t\left( \frac{p_i(\bm{y}_i^{(j_{+})}|\bm{\theta})}{p_i(\bm{y}_i|\bm{\theta})} \right) \Bigg\}\Bigg),
\]
and $g(q_*)$ is a constant depending on $q_*$ but not on $\bm{\theta}$. 
\end{theorem}

An empirical estimator of the population function, $d_{\rm GSM}(q_*,p_*)$, is given by
\begin{align}\label{discrete empirical function multi}
\hat{d}_{\rm GSM}(\bm{\theta})=\frac{1}{n}\sum_{i=1}^n \rho_i^{\rm GSM}(\bm{y}_i|\bm{\theta}),
\end{align}
where
\begin{align*}
\rho_i^{\rm GSM}(\bm{y}_i|\bm{\theta})=\sum_{j=1}^d t\left( \frac{p_i(\bm{y}_i^{(j_{+})}|\bm{\theta})}{p_i(\bm{y}_i|\bm{\theta})} \right)^2 + t\left( \frac{p_i(\bm{y}_i|\bm{\theta})}{p_i(\bm{y}_i^{(j_{-})}|\bm{\theta})}\right)^2-2t\left( \frac{p_i(\bm{y}_i^{(j_{+})}|\bm{\theta})}{p_i(\bm{y}_i|\bm{\theta})} \right).
\end{align*}
The generalized score matching estimator for $\bm{\theta}$ is then defined as
\[
\hat{\bm{\theta}}=\argmin_{\bm{\theta}}\hat{d}_{\rm GSM}(\bm{\theta}).
\]

The following theorem states the local consistency.
\begin{theorem}
Assume that the model $p(\bm{y}|\bm{x}_i,\bm{\theta})$ is correct, that is, $p_i(\bm{y}_i|\bm{\theta}_0)=q(\bm{y}|\bm{x}_i)$ for $i\in\{1,\ldots,n\}$ and further suppose that the model is identifiable, i.e. for each $\bm{\theta}\neq\bm{\theta}_0$, there exists a set of $\bm{y}$ of positive probability under $p_i(\bm{y}_i|\bm{\theta}_0)$ such that $p(\bm{y}|\bm{x}_i,\bm{\theta})\neq p_i(\bm{y}_i|\bm{\theta}_0)$. Then, $D_{\rm GSM}(q_*,p_*)=0$ if and only if $\bm{\theta}=\bm{\theta}_0$, where $D_{\rm GSM}$ is defined in (\ref{discrete population function multi}).
\end{theorem}

\section{Derivatives of Score Matching Objective Function}\label{sup sec::score hessian}
In this section, we will present derivatives of proposed score matching objective functions for the truncated Gaussian regression model and the CMP regression model.

\subsection{Asymptotic Variance for Truncated Gaussian Regression Models}\label{sup sec:: asy var tgr}
In this section, we provide detailed derivatives of the score matching objective function for truncated Gaussian regression models.

The score vector is given by
\[
\frac{\partial \hat{d}_{\rm SM}(\bm{\theta})}{\partial\bm{\theta}}=\begin{pmatrix} \bm{s}_{\bm{B}}\\\bm{s}_{\bm{\Lambda}} \end{pmatrix},
\]
where
\begin{align*}
\bm{s}_{\bm{B}}&=\frac{\partial \hat{d}_{\rm SM}(\bm{\theta})}{\partial{\rm vec}(\bm{B})}\\
&=\frac{1}{n}\sum_{i=1}^n {\rm vec}\left\{  4\bm{\Lambda}\bm{t}(\widetilde{\bm{y}}_i)\bm{x}_i^\top-2\bm{\Lambda}\bm{T}(\widetilde{\bm{y}}_i) \bm{T}(\widetilde{\bm{y}}_i)  \bm{\Lambda} (\bm{t}(\widetilde{\bm{y}}_i)-\bm{B}\bm{x}_i) \bm{x}_i^\top     \right\},
\end{align*}
and
\begin{align*}
\bm{s}_{\bm{\Lambda}}=&\frac{\partial \hat{d}_{\rm SM}(\bm{\theta})}{\partial{\rm vech}(\bm{\Lambda})}\\
=&\frac{1}{n}\sum_{i=1}^n \bm{D}_p^\top {\rm vec}\Big\{  -4 (\bm{t}(\widetilde{\bm{y}}_i)-\bm{B}\bm{x}_i)\bm{t}(\widetilde{\bm{y}}_i)^\top -2\bm{T}(\widetilde{\bm{y}}_i) \bm{T}(\widetilde{\bm{y}}_i) \\
&+2\bm{T}(\widetilde{\bm{y}}_i) \bm{T}(\widetilde{\bm{y}}_i) \bm{\Lambda} (\bm{t}(\widetilde{\bm{y}}_i)-\bm{B}\bm{x}_i)  (\bm{t}(\widetilde{\bm{y}}_i)-\bm{B}\bm{x}_i)^\top  \Big\} ,
\end{align*}
where $\bm{D}_p\in\mathbb{R}^{p^2\times p(p+1)/2}$ is the duplication matrix. The Hessian matrix is
\[
-\frac{\partial^2\hat{d}_{\rm SM}(\bm{\theta})}{\partial\bm{\theta}\partial\bm{\theta}^\top}=\begin{pmatrix} \mathcal{I}_{\bm{B}\bm{B}} & \mathcal{I}_{\bm{B}  \bm{\Lambda}}\\\mathcal{I}_{\bm{\Lambda}\bm{B}} &\mathcal{I}_{\bm{\Lambda}\bm{\Lambda}} \end{pmatrix},
\]
where
\begin{align*}
\mathcal{I}_{\bm{B}\bm{B}}&=-\frac{1}{n}\sum_{i=1}^n 2(\bm{x}_i\bm{x}_i^\top)\otimes (\bm{\Lambda}\bm{T}(\widetilde{\bm{y}}_i) \bm{T}(\widetilde{\bm{y}}_i)  \bm{\Lambda}),
\end{align*}

\begin{align*}
\mathcal{I}_{\bm{B}  \bm{\Lambda}}=&\mathcal{I}_{\bm{\Lambda}\bm{B}}^\top\\
=&-\frac{1}{n}\sum_{i=1}^n \Big\{  (4\bm{x}_i \bm{t}(\widetilde{\bm{y}}_i)^\top)\otimes \bm{I}_p - [\bm{x}_i(\bm{t}(\widetilde{\bm{y}}_i)-\bm{B}\bm{x}_i)^\top]\otimes (2\bm{\Lambda}\bm{T}(\widetilde{\bm{y}}_i) \bm{T}(\widetilde{\bm{y}}_i) )  \\ 
&-2(\bm{x}_i(\bm{t}(\widetilde{\bm{y}}_i)-\bm{B}\bm{x}_i)^\top\bm{\Lambda} \bm{T}(\widetilde{\bm{y}}_i) \bm{T}(\widetilde{\bm{y}}_i)  )\otimes \bm{I}_p\Big\}\bm{D}_p,
\end{align*}
and
\begin{align*}
\mathcal{I}_{\bm{\Lambda}\bm{\Lambda}}=&-\frac{1}{n}\sum_{i=1}^n  2\bm{D}_p^\top \left( \left[ (\bm{t}(\widetilde{\bm{y}}_i)-\bm{B}\bm{x}_i)  (\bm{t}(\widetilde{\bm{y}}_i)-\bm{B}\bm{x}_i)^\top \right] \otimes \left[ \bm{T}(\widetilde{\bm{y}}_i) \bm{T}(\widetilde{\bm{y}}_i)  \right] \right)  \bm{D}_p,
\end{align*}
with $\otimes$ being the Kronecker product.

\subsection{Asymptotic Variance for CMP Regression Models}\label{sup sec:: asy var cmp}
In this section, we provide detailed derivatives of the generalized score matching objective function for CMP regression models.

The score vector is given by
\[
\frac{\partial \hat{d}_{\rm GSM}(\bm{\theta})}{\partial\bm{\theta}}=\begin{pmatrix} \bm{s}_{\bm{\beta}}\\s_\nu \end{pmatrix},
\]
where
\begin{align*}
\bm{s}_{\bm{\beta}}&=\frac{\partial \hat{d}_{\rm GSM}(\bm{\theta})}{\partial\bm{\beta}}=\frac{1}{n}\sum_{i=1}^n\frac{\partial \rho_{\rm GSM}(\bm{\theta})}{\partial\lambda_i}\frac{d\lambda_i}{d(\bm{x}_i^\top\bm{\beta})}\frac{\partial \bm{x}_i^\top\bm{\beta}}{\partial \bm{\beta}}\\
&=\frac{1}{n}\sum_{i=1}^n \frac{\partial \rho_{\rm GSM}(\bm{\theta})}{\partial\lambda_i}\lambda_i \bm{x}_i,
\end{align*}
with
\[
\frac{\partial \rho_{\rm GSM}(\bm{\theta})}{\partial\lambda_i}= -\frac{2t\left(\frac{\lambda_i}{(y_i+1)^\nu}\right)^3}{(y_i+1)^\nu}-\frac{2t\left(\frac{\lambda_i}{y_i^\nu}\right)^3}{y_i^\nu}+\frac{2t\left(\frac{\lambda_i}{(y_i+1)^\nu}\right)^2}{(y_i+1)^\nu},
\]
and
\begin{align*}
s_\nu&=\frac{\partial \hat{d}_{\rm GSM}(\bm{\theta})}{\partial\nu}=\frac{1}{n}\sum_{i=1}^n\frac{\partial \rho_{\rm GSM}(\bm{\theta})}{\partial\nu}\\
&=\frac{1}{n}\sum_{i=1}^n \Bigg\{  \frac{2\lambda_i\ln (y_i+1)}{(y_i+1)^\nu}t\left(\frac{\lambda_i}{(y_i+1)^\nu}\right)^3+\frac{2\lambda_i\ln (y_i)}{y_i^\nu}t\left(\frac{\lambda_i}{y_i^\nu}\right)^3-\frac{2\lambda_i\ln (y_i+1)}{(y_i+1)^\nu}t\left(\frac{\lambda_i}{(y_i+1)^\nu}\right)^2 \Bigg\}.
\end{align*}
The Hessian matrix is
\[
-\frac{\partial^2\hat{d}_{\rm GSM}(\bm{\theta})}{\partial\bm{\theta}\partial\bm{\theta}^\top}=\begin{pmatrix} \mathcal{I}_{\bm{\beta}\bm{\beta}} & \mathcal{I}_{\bm{\beta}\nu}\\\mathcal{I}_{\nu\bm{\beta}} &\mathcal{I}_{\nu\nu} \end{pmatrix},
\]
where
\begin{align*}
\mathcal{I}_{\bm{\beta}\bm{\beta}}&=-\frac{1}{n}\sum_{i=1}^n \left(\frac{\partial^2 \rho_{\rm GSM}(\bm{\theta})}{\partial\lambda_i^2} \frac{d\lambda_i}{d(\bm{x}_i^\top\bm{\beta})} + \frac{\partial \rho_{\rm GSM}(\bm{\theta})}{\partial\lambda_i}\frac{\partial}{\partial \lambda_i} \frac{d\lambda_i}{d(\bm{x}_i^\top\bm{\beta})} \right)\frac{d\lambda_i}{d(\bm{x}_i^\top\bm{\beta})}\bm{x}_i\bm{x}_i^\top\\
&=-\frac{1}{n}\sum_{i=1}^n \left(\frac{\partial^2 \rho_{\rm GSM}(\bm{\theta})}{\partial\lambda_i^2} \lambda_i + \frac{\partial \rho_{\rm GSM}(\bm{\theta})}{\partial\lambda_i} \right)\lambda_i\bm{x}_i\bm{x}_i^\top,
\end{align*}
with
\begin{align*}
\frac{\partial^2 \rho_{\rm GSM}(\bm{\theta})}{\partial\lambda_i^2}=\frac{6t\left(\frac{\lambda_i}{(y_i+1)^\nu}\right)^4}{(y_i+1)^{2\nu}}+\frac{6t\left(\frac{\lambda_i}{y_i^\nu}\right)^4}{y_i^{2\nu}}-\frac{4t\left(\frac{\lambda_i}{(y_i+1)^\nu}\right)^3}{(y_i+1)^{2\nu}},
\end{align*}
\begin{align*}
\mathcal{I}_{\bm{\beta}\nu}=&\mathcal{I}_{\nu\bm{\beta}}^\top=-\frac{1}{n}\sum_{i=1}^n \frac{\partial^2 \rho_{\rm GSM}(\bm{\theta})}{\partial\lambda_i\partial \nu}\lambda_i \bm{x}_i\\
=&-\frac{1}{n}\sum_{i=1}^n \Bigg\{ \frac{2\ln(y_i+1)}{(y_i+1)^\nu}t\left(\frac{\lambda_i}{(y_i+1)^\nu}\right)^3\left[1-\frac{3\lambda_i}{\lambda_i+(y_i+1)^\nu}\right] + \frac{2\ln(y_i)}{y_i^\nu}t\left(\frac{\lambda_i}{y_i^\nu}\right)^3\left[1-\frac{3\lambda_i}{\lambda_i+y_i^\nu}\right]\\
 &+\frac{2\ln(y_i+1)}{(y_i+1)^\nu}t\left(\frac{\lambda_i}{(y_i+1)^\nu}\right)^2\left[-1+\frac{2\lambda_i}{\lambda_i+(y_i+1)^\nu}\right]  \Bigg\} \lambda_i \bm{x}_i,
\end{align*}

and
\begin{align*}
\mathcal{I}_{\nu\nu}=&-\frac{1}{n}\sum_{i=1}^n\Bigg\{ \frac{2\lambda_i[\ln(y_i+1)]^2}{(y_i+1)^\nu}t\left(\frac{\lambda_i}{(y_i+1)^\nu}\right)^3\left[-1+\frac{3\lambda_i}{\lambda_i+(y_i+1)^\nu}\right] \\
&+ \frac{2\lambda_i[\ln(y_i)]^2}{y_i^\nu}t\left(\frac{\lambda_i}{y_i^\nu}\right)^3\left[-1+\frac{3\lambda_i}{\lambda_i+y_i^\nu}\right]\\
&+\frac{2\lambda_i[\ln(y_i+1)]^2}{(y_i+1)^\nu}t\left(\frac{\lambda_i}{(y_i+1)^\nu}\right)^2\left[1-\frac{2\lambda_i}{\lambda_i+(y_i+1)^\nu}\right]\Bigg\}.
\end{align*}

\bibliographystyle{99}

\end{document}